\documentclass[11pt,article]{elsarticle}

\usepackage{lineno,hyperref}

\usepackage[lmargin=.75in,rmargin=.75in,tmargin=1.in,bmargin=1in]{geometry}
\usepackage{amsmath}
\usepackage{amsthm}
\usepackage{tabulary}
\usepackage{booktabs}
\usepackage{amsfonts}
\usepackage{amssymb}
\usepackage{graphicx}
\usepackage{multicol}
\usepackage{wrapfig}
\usepackage{setspace}
\usepackage{epstopdf}
\usepackage{graphicx}
\usepackage{caption}
\usepackage{subcaption}
\usepackage{color}
\usepackage{algorithm}
\usepackage{algpseudocode}
\newtheorem*{remark}{Remark}

\journal{arXiv}

\begin{document}

\begin{frontmatter}

\title{Ensemble-marginalized Kalman filter for linear time-dependent PDEs with noisy boundary conditions: Application to heat transfer in building walls}

\author[NU]{Marco Iglesias}
\ead{marco.iglesias@nottingham.ac.uk}
\author[KAUST]{Zaid Sawlan \corref{corrauthor}}
\cortext[corrauthor]{Corresponding author}
\ead{zaid.sawlan@kaust.edu.sa}
\author[KAUST,Ur]{Marco Scavino}
\ead{marco.scavino@kaust.edu.sa}
\author[KAUST]{Ra\'ul Tempone}
\ead{raul.tempone@kaust.edu.sa}
\author[NUA]{Christopher Wood}
\ead{christopher.wood@nottingham.ac.uk}

\address[NU]{School of Mathematical Sciences, University of Nottingham, Nottingham, UK}
\address[KAUST]{King Abdullah University of Science and Technology (KAUST), Computer, Electrical and Mathematical Science and Engineering Division (CEMSE), Thuwal 23955-6900, Saudi Arabia}
\address[NUA]{Department of Architecture and Built Environment, University of Nottingham, Nottingham, UK}
\address[Ur]{Instituto de Estad\'{\i}stica (IESTA), Universidad de la Rep\'ublica, Montevideo, Uruguay}

\begin{abstract}
In this work, we present the ensemble-marginalized Kalman filter (EnMKF), a sequential algorithm analogous to our previously proposed approach \cite{sawlan1, sawlan2}, for estimating the state and parameters of linear parabolic partial differential equations in initial-boundary value problems when the boundary data are noisy. We apply EnMKF to infer the thermal properties of building walls and to estimate the corresponding heat flux from real and synthetic data. Compared with a modified Ensemble Kalman Filter (EnKF) that is not marginalized, EnMKF reduces the bias error, avoids the collapse of the ensemble without needing to add inflation, and converges to the mean field posterior using $50\%$ or less of the ensemble size required by EnKF. According to our results, the marginalization technique in EnMKF is key to performance improvement with smaller ensembles at any fixed time.
\end{abstract}



\end{frontmatter}


\begin{table}[htbp]\caption{List of notation}
\begin{center}
\begin{tabular}{r p{15cm} }
\toprule
$\mathop{\theta}\limits_{p\times 1}$   & 		parameter vector \\
$\mathop{A_{\theta}}\limits_{n\times n}$     &          forward model matrix operator \\
$\mathop{T_k}\limits_{n\times 1}$ 	& 		state vector at time $t_k$ \\
$\mathop{B_{\theta}}\limits_{n \times \ell}$     &          control matrix operator \\
$\mathop{u_k}\limits_{\ell \times 1}$ 	& 		control vector at time $t_k$ \\
$\mathop{F}\limits_{\ell \times \ell}$ & 		control evolution matrix operator \\
$\mathop{H}\limits_{m \times n}$     &          observation matrix operator \\
$\mathop{y_k}\limits_{m\times 1}$ 	& 		observation vector at time $t_k$  \\
$y_{1:k} = (y_1, \ldots, y_k)  $ 	& 		observation vectors from time $t_1$ to $t_k$  \\
$\mathop{z_k}\limits_{\ell \times 1}$ 	& 		observed control vector at time $t_k$  \\
$z_{1:k} = (z_1, \ldots, z_k)  $ 	& 		observed control vectors from time $t_1$ to $t_k$  \\
$\mathop{X_k}\limits_{(n+p)\times 1}$  & 		augmented state-parameter vector at time $t_k$\\
$w_k$		& centered Gaussian noise of the state vector $T_k$ with covariance matrix $W_k$ \\
$v_k$		& centered Gaussian noise of the observation vector $y_k$ with covariance matrix $V_k$ \\
$q_k$		& centered Gaussian noise of the control vector $u_k$ with covariance matrix $Q_k$ \\
$c_k$ 		&     centered Gaussian noise of the observed control vector $z_k$ with covariance matrix $C_k$ \\
$T_{k|k}$ 	& 	$E[ T_k | \theta, z_{1:k}, y_{1:k} ]$ \\
$T_{k|k-1}$ 	& 	$E[ T_k | \theta, z_{1:k}, y_{1:k-1} ]$ \\
$P_{k|k}$ 	& 	$Cov(T_k | \theta, z_{1:k}, y_{1:k})$ \\
$P_{k|k-1}$ 	& 	$Cov(T_k | \theta, z_{1:k}, y_{1:k-1})$ \\
$u_{k|k}$ 	& 	$E[ u_k | z_{1:k}]$ \\
$P^{u}_{k|k}$ 	& 	$Cov(u_k | z_{1:k})$ \\
$\mathcal{P}_{k|k-1}$	& 	$Cov(X_{k}|z_{1:k}, y_{1:k-1})$ \\
\bottomrule
\end{tabular}
\end{center}
\label{tab:TableOfNotationForMyResearch}
\end{table}

\section{Introduction}

Uncertainty in the thermophysical properties of building envelopes has been recognized as one of the main sources of uncertainty when simulating the thermal performance of buildings \cite{li2014solid,cesaratto2013measuring,doran2000detr,asdrubali2014evaluating,hong2006impact}. Simulations of this kind are essential to the development of cost-effective retrofitting measures aimed at improving the energy efficiency of existing buildings. Moreover, accurate uncertainty analyses of the energy performance of the housing stock are vital for drafting policies to reduce global carbon emissions \cite{biddulph2014inferring,li2014solid}. Most existing approaches use simplified heat diffusion models to quantify uncertainty in the thermophysical properties of walls \cite{biddulph2014inferring,gori2017inferring}. However, recent works have suggested that a PDE-constrained data-assimilation approach is required to accurately estimate the thermophysical properties of walls \cite{REnKA,sawlan2}.

Most existing statistical approaches that infer thermophysical properties of walls \cite{biddulph2014inferring,gori2017inferring,REnKA} assume that the boundary temperatures, $T_{int}(t)$ and $T_{ext}(t)$, can be approximated by their corresponding measurements recorded on the internal and external surfaces of a wall by temperature sensors. Although these approaches ignore the uncertainty in those measurements, they enable the construction of a well-defined parameter-to-measurements map that can be inverted via existing frameworks \cite{kaipio2005statistical}. However, we recently showed that ignoring the uncertainty in boundary temperatures can lead to biased estimates of the thermophysical properties and, thus, reduce the accuracy of uncertainty analyses in energy-performance evaluations of buildings \cite{sawlan2}. In order to account for the uncertainty of boundary measurements within a PDE-constrained Bayesian framework, we developed a marginalization technique that assimilates the available data all at once. This approach was both (i) accurate in identifying thermophysical properties of walls, and (ii) useful for developing off-line experimental design strategies to reduce the measurement campaigns. 

Following these recent advances, practitioners would substantially benefit from a data assimilation framework capable of updating the probabilistic knowledge of thermophysical properties as new measurements become available. Such a framework could be used to assess in real time whether a specific level of accuracy or uncertainty has been achieved, in order to reduce the long measurement campaigns that often involve several weeks \cite{doran2000detr}. The objective of this work is to develop a sequential Bayesian approach to infer the thermal properties of walls and estimate the corresponding heat flux within a computationally tractable framework that accounts for the uncertainty in the measurements of the boundary temperatures. 

Bayesian filtering \cite{Doucet2000,asch2016data} can be used to compute the desired distribution for a joint state-parameter assimilation problem. However, sampling is required to approximate the desired distribution when the Bayesian posterior cannot be expressed in a closed form. For example, particle filters use importance sampling to represent the posterior distribution. Unfortunately, particle filters may become computationally prohibitive \cite{Failure} and, thus, unsuitable for practical implementation in real-time data assimilation. As an alternative to fully Bayesian samplers, the ensemble Kalman filter (EnKF) is the preferred method for data assimilation in partially observed systems where the underlying dynamics are nonlinear and the state space is high-dimensional \cite{evensenENKF1}. Initially proposed for state estimation \cite{burgersEnKF}, modified versions of EnKF have been proposed to address joint state-parameter estimation \cite{evensenENKF2}. The EnKF methodology uses a Gaussian approximation for the predictive step in the Bayesian filtering framework; this enables the characterization of the Bayesian posterior within the analysis step as a Gaussian distribution with mean and covariance computed via standard Kalman-like formulas. Different approaches to characterize this Gaussian distribution have led to several EnKF algorithms including the stochastic EnKF and the ensemble square-root Kalman filter that are reviewed elsewhere \cite{law2015data, evensenENKF1}. 

The Gaussian approximations within the EnKF framework do not lead to algorithms with asymptotic convergence to the desired posterior when the underlying dynamics of the state-parameter system are nonlinear \cite{doi:10.1137/140965363}. Nevertheless, EnKF has been successfully used in numerous applications, including oceanography, meteorology, subsurface flow and engineering \cite{understandEnKF,iglesias}. In most of these studies, the success of EnKF was reliant on ad-hoc fixes such as covariance inflation and localization \cite{inflation1, inflation2,understandEnKF}. More recently, regularized versions of EnKF were proposed that merged ideas from particle filters with iterative regularization techniques \cite{SMC_REnKA,EnsembleYO,REnKA}. For example, a regularizing EnKF was applied for parameter estimation of thermal properties of walls \cite{REnKA}. This approach, however, focused entirely on the Bayesian parameter estimation of the thermophysical properties and did not consider joint state-parameter estimation. Consequently, the resulting algorithm required that the heat transfer model was run from the initial start time until the current assimilation time to update the state of the system (i.e., the wall-temperature profile). A joint state-parameter estimation approach for the real-time assimilation of in-situ measurements that enabled sequential updating of both the parameters and the state would be much more efficient.

In this paper, we study the joint state-parameter estimation problem within a setting of linear PDEs with noisy boundary conditions motivated by the aforementioned application which involves the mathematical modeling of heat transfer through walls. More specifically, we are interested in the real-time uncertainty quantification of thermophysical properties of walls via the sequential assimilation of in-situ measurements of the wall's surface temperature and heat flux. We propose an ensemble-marginalized Kalman filter (EnMKF) to approximate the joint state-parameter Bayesian estimation problem for a partially observed linear system that is obtained from the discretization of linear time-dependent PDEs. In this general setting, we obtain a linear forward operator that depends nonlinearly on a vector of parameters while the observation operator is linear and independent of all parameters. In our specific application, we also adapt EnMKF and EnKF to handle an observation operator that depends linearly on the thermal resistance parameter. 

The remainder of the paper is organized as follows. In Section \ref{section1}, we introduce the state-space formulation that can be used to analyze a class of linear PDE problems. We present the filtering approach for estimating the unknown state-parameter vectors in Section \ref{section2}. In particular, subsection \ref{MBF} provides a Bayesian formulation for the proposed marginalized filtering algorithms. Subsection \ref{EnMKF} contains our novel ensemble-marginalized Kalman filter (EnMKF). In subsection \ref{EnKF}, we consider a slightly modified EnKF where the boundary conditions are sampled from their filtered distribution, and we describe the primary differences between EnMKF and the modified EnKF. The performances of the EnMKF and the modified EnKF algorithms are compared using real and synthetic data in Section \ref{application}. Conclusions are summarized in Section \ref{conc}.


\section{General formulation}
\label{section1}

Consider the linear time-dependent parabolic initial-boundary value problem on the domain $D\times [0,t_{N}] \subset \mathbb{R}^{d+1}$:

\begin{equation}
\begin{cases}
\label{gmodel}
\frac{\partial T}{\partial t} +  L_{\theta} T = 0,\,  x \in D,\,  t \in [0,t_{N}] \\
T(x,t) = u(x,t),\,  x \in \partial D,\, t \in [0,t_{N}] \\
T(x,0) = T_0(x),\, x \in D, \\
\end{cases}
\end{equation}
where $L_{\theta} T = - \sum_{i,j=1}^{d} a_{ij}(x) \frac{\partial^2 T}{\partial x_i \partial x_j} + \sum_{i=1}^{d} b_{i}(x) \frac{\partial T}{\partial x_i} + c(x) T$ and $\theta = (a_{11}, \ldots, a_{dd}, b_1, \ldots, b_d, c)$. Our goal is to estimate $\theta$ and $T$ sequentially in time given partial observations of $T(x,t)$ and measurements of $u(x,t)$. As a special case, we consider in Section \ref{application} the heat equation to model the heat transfer process through building walls and, given real measurements, we estimate the thermal properties of a brick wall.


By discretizing \eqref{gmodel} using finite differences or finite elements \cite{sawlan1}, we can derive the discrete state-space model 


\begin{equation}
\begin{cases}
T_{k} = A_{\theta} T_{k-1} + B_{\theta} u_{k} + w_k, \\
y_{k} = H T_{k} + v_k, 
\end{cases}
\label{system1}
\end{equation}
where $\mathop{T_k}\limits_{n\times 1}$ is the state vector at time $t_k$, $\mathop{y_k}\limits_{m\times 1}$ corresponds to measurements of $T_k$, $A_{\theta}$ is the forward model operator that depends on the parameter vector $\mathop{\theta}\limits_{p\times 1}$, $H$ is a linear observational map, $B_{\theta}$ is the control matrix operator, and we assume that the control vector $\mathop{u_k}\limits_{\ell \times 1}$ follows another state observation system:


\begin{equation}
\begin{cases}
u_{k} = F u_{k-1} + q_k,\\
z_{k} = u_{k} + c_k,
\end{cases}
\label{system2}
\end{equation}
where $F$ is a user-defined linear evolution operator (see subsection \ref{KF_sec}) and $\mathop{z_k}\limits_{\ell \times 1}$ is the observed control vector at time $t_k$. We further assume that $w_k, v_k, q_k, c_k$ are independent centered Gaussian random vectors with covariances $W_k, V_k, Q_k, C_k$, respectively. 


The joint Bayesian estimation of $\theta$ and $T_k$, in a sequential filtering approach, as the observations $\{z_{1:k}, y_{1:k}\}$ become available can be now posed in terms of the computation of the conditional density $p(T_{k} , \theta | z_{1:k}, y_{1:k} )$. Following the marginalization technique \cite{sawlan1,sawlan2} and ideas from marginalized particle filters \cite{marginalizedPF,doucet2000rao,vsmidl2011marginalized}, we exploit the linear structure of \eqref{system1} and propose an ensemble-marginalized Kalman filter (EnMKF) that provides an approximation of $p(T_{k} , \theta \vert  z_{1:k}, y_{1:k} )$. From \eqref{system1}, it is easy to approximate the conditional distribution $p(T_{k} , \theta \vert  u_k, y_{1:k} )$. In order to incorporate the error of the boundary conditions, we use the tractability of the distribution $p(u_{k}\vert z_{1:k})$. In other words, by noticing from \eqref{system2} that $u_k$ follows a linear dynamic system independent of (\ref{system1}), and under further Gaussian assumptions on $u_0$, it follows that the distribution $p(u_{k}\vert  z_{1:k})$ can be expressed in closed form by using the Kalman filter \cite{Kalman}. The desired joint distribution $p(T_{k} , \theta \vert  z_{1:k}, y_{1:k} )$ is then approximated by integrating out $u_k$ from the fully joint distribution $ p(T_{k} , \theta, u_k \vert  z_{1:k}, y_{1:k} ) = p(T_{k} , \theta \vert  u_k, y_{1:k} ) p(u_{k}\vert  z_{1:k})$.

\section{Methodology}
\label{section2}

In this section, we introduce the ensemble-marginalized Kalman filter (EnMKF) as a state-parameter estimation algorithm for the partially observed linear system \eqref{system1}. The boundary conditions $u_k$ are estimated first by applying the classical Kalman filter to \eqref{system2}. Next, we derive a marginalized Kalman filter for the state $T_k$, assuming that $\theta$ is known. Then, this intermediate result is generalized to develop a joint filter for the state $T_k$ and parameter $\theta$ based on the EnKF framework. EnMKF differs from EnKF in the prediction step; in EnMKF, the prediction covariance of $T_k$ is inflated by the expectation of the analysis covariance matrix of $u_k$ with respect to $\theta$.

\subsection{Kalman filtering for $u_{k}$}\label{KF_sec}

The state-space model for $u_k$ in \eqref{system2} is linear and independent from $T_k$. Therefore, we can implement Kalman filter given the observation $z_k$ as follows:

\begin{enumerate}
\item Prediction step:
\begin{equation}
\begin{cases}
u_{k|k-1} =   F u_{k-1|k-1}, \\
P^{u}_{k|k-1} = F P^{u}_{k-1|k-1} F' + Q_k,
\end{cases}
\label{kalmanpredstep}
\end{equation}
\item Analysis step: 
\begin{equation}
\begin{cases}
K^{u}_k = P^{u}_{k|k-1} \left( P^{u}_{k|k-1} + C_{k} \right)^{-1},\\
u_{k|k} = u_{k|k-1} + K_{k} \left(z_{k} - u_{k|k-1} \right),\\
P^{u}_{k|k} = \left( I -  K_{k} \right) P^{u}_{k|k-1},
\end{cases}
\label{kalmananalysis}
\end{equation}
where $K^u_k$ is the Kalman gain matrix of the state-space model for the control vector $u_k$.
\end{enumerate}

Assuming $u_0$ is Gaussian, the Kalman filter provides the exact filtering distribution as $N(u_{k|k},P^{u}_{k|k})$.

In many applications, the forward operator $F$ is unknown and autoregressive models must be considered \cite{autoregressive}. Some proposals for the autoregressive forward models are: 

\begin{itemize}
\item Random walk model AR(1)
\begin{equation*}
u_{k} =  u_{k-1} + q_{k},
\end{equation*}

\item Random increment model AR(2)
\begin{equation*}
u_{k} = 2 u_{k-1} - u_{k-2} + q_{k},
\end{equation*}

\end{itemize}
with $q_{k} \sim N(0, Q_k)\,.$

In subsection \ref{results1}, we impose AR(1) and AR(2) on the boundary temperatures, $T_{int}(t)$ and $T_{ext}(t)$, and compare the filtered results with the real data.

\begin{remark}
Although the state-space model for $u_k$ is independent from $T_k$, we apply the Kalman filter within the time propagation of either EnMKF or EnKF. 
\end{remark}

\subsection{Marginalized Kalman filter for $T_{k}|\theta$}
\label{margkalman}
Before deriving the filtering algorithm for the joint state-parameter vector, we derive the marginalized Kalman filter for the conditional state $T_k|\theta$ given the observations $(z_{1:k}, y_{1:k})$. Assuming $T_0$ is Gaussian, the Kalman filter is sufficient to estimate the filtering distribution of $T_k$ given $\theta$. We derive the conditional mean and the conditional covariance for $T_{k}|\theta$ given the observations $(z_{1:k}, y_{1:k-1})$:
\begin{eqnarray}
T_{k|k-1} &=& E[ T_{k} | \theta, z_{1:k}, y_{1:k-1} ]  =  E[ A_{\theta} T_{k-1} + B_{\theta} u_{k} + w_k| \theta, z_{1:k}, y_{1:k-1} ] \nonumber \\
&=& A_{\theta}  E[ T_{k-1} | \theta, z_{1:k-1}, y_{1:k-1} ] + B_{\theta} E [u_{k}| z_{1:k} ] \nonumber \\
&=& A_{\theta}  T_{k-1|k-1} + B_{\theta} u_{k|k} \, .  \label{margmean} \\
P_{k|k-1}  &=&  Cov[  T_{k} | \theta, z_{1:k}, y_{1:k-1} ] \nonumber \\
&=&  Cov[ A_{\theta} T_{k-1} + B_{\theta} u_{k} + w_k | \theta, z_{1:k}, y_{1:k-1} ] \nonumber \\
&& \text{(using the mutual independence of $T_{k-1}, u_{k}$ and $w_k$)} \nonumber \\
&=&  Cov[ A_{\theta} T_{k-1} | \theta, z_{1:k-1}, y_{1:k-1} ] + Cov[ B_{\theta} u_{k} | \theta, z_{1:k} ] +  Cov[ w_k ] \nonumber \\
&=& A_{\theta} P_{k-1|k-1} A'_{\theta} + B_{\theta} P^{u}_{k|k} B'_{\theta} + W_{k}. \label{margcov}
\end{eqnarray}

From \eqref{margmean} and \eqref{margcov}, the marginalized Kalman filter for $T_{k}|\theta$ is summarized by
\begin{enumerate}
\item Prediction step:
\begin{eqnarray*}
T_{k|k-1} &=& A_{\theta} T_{k-1|k-1} + B_{\theta} u_{k|k}, \\
P_{k|k-1} &=&  A_{\theta} P_{k-1|k-1} A_{\theta}'+ B_{\theta} P^{u}_{k|k} B_{\theta}' + W_{k},
\end{eqnarray*}
where $T_{k|k-1}$ is the state prediction at time $k$, given $\theta$ and the observations $(z_{1:k}, y_{1:k-1})$, and $P_{k|k-1}$ is the corresponding covariance matrix.

\item Analysis step:
\begin{eqnarray*}
K_k &=& P_{k|k-1}  H' \left(  H  P_{k|k-1}  H' + V_{k} \right)^{-1},\\
T_{k|k} &=& T_{k|k-1} + K_{k} \left(y_{k} - H T_{k|k-1} \right),\\
P_{k|k} &=& \left( I -  K_{k}  H \right) P_{k|k-1},
\end{eqnarray*}
where $T_{k|k}$ is the state estimation at time $k$, given $\theta$ and the observations $(z_{1:k}, y_{1:k})$, and $K_{k}$ is the Kalman gain.
\end{enumerate}

\begin{remark}
The prediction step can be generalized under the Bayesian approach and corresponds to the following equation:
\begin{eqnarray*}
p(T_{k} | \theta, z_{1:k}, y_{1:k-1} ) & =& \int p(T_{k} | \theta,z_{1:k}, T_{k-1} ) p(T_{k-1} | \theta, z_{1:k-1}, y_{1:k-1} )\, d T_{k-1} \\
& =& \int \left( \int p(T_{k} | \theta, u_{k}, T_{k-1} ) p(u_{k}| z_{1:k}) \, du_{k} \right) p(T_{k-1} | \theta, z_{1:k-1}, y_{1:k-1} )\, d T_{k-1} \,,
\end{eqnarray*}
where the marginalization of $u_k$ appears explicitly. The analysis step is generalized under the Bayesian approach as
\begin{eqnarray*}
p(T_{k} | \theta, z_{1:k}, y_{1:k} ) &=&  \frac{p(y_{k} | T_{k}) p(T_{k} | \theta, z_{1:k}, y_{1:k-1} )}{\int p(y_{k} | T_{k}) p(T_{k} | \theta, z_{1:k}, y_{1:k-1} ) d T_{k}} \\ 
\\
&\propto& p(y_{k} | T_{k}) p(T_{k} | \theta, z_{1:k}, y_{1:k-1} ) \,.
\end{eqnarray*}
\end{remark}

\subsection{Marginalized Bayesian filtering for $T_k$ and $\theta$}
\label{MBF}

We now derive filtering algorithms to estimate the joint state-parameter vector, $X_k = \left[\, \theta \,\,\,\,\, T_k \, \right]'$, given the observations $(z_{1:k}, y_{1:k})$ up to time $k$. In this case, the evolution of the joint vector, $X_k$, is nonlinear and approximate filtering algorithms are needed. We derive first the underlying Bayesian formulation for the filtering algorithms. The static parameters, $\theta$, have a trivial evolution in the prediction step and therefore we neglect their time subscript.


\begin{enumerate}
\item Initialization:
\begin{eqnarray*}
\theta \sim \rho_{\Theta}(\theta), \\
T_{0} \sim \rho_{T_0}( T_0),
\end{eqnarray*}
where $\rho_{\Theta}(\cdot)$ and $\rho_{T_{0}}(\cdot)$ are the prior distributions of $\theta$ and $T_0$, respectively.

\item Prediction step:
\begin{eqnarray*}
p(X_{k}| z_{1:k}, y_{1:k-1} ) &=& p(T_{k}, \theta | z_{1:k}, y_{1:k-1} ) \\
&=& p(T_{k}| \theta, z_{1:k}, y_{1:k-1} ) p(\theta | z_{1:k}, y_{1:k-1}) \\
& =& \left( \int p(T_{k} | \theta, z_{1:k}, T_{k-1} ) p(T_{k-1} | \theta, z_{1:k-1}, y_{1:k-1} )\, d T_{k-1} \right)  p(\theta | z_{1:k-1}, y_{1:k-1}) \\
& =& \int p(T_{k} | \theta, z_{1:k}, T_{k-1} ) p(T_{k-1}, \theta | z_{1:k-1}, y_{1:k-1} ) \, d T_{k-1} \\
& =& \int p(T_{k} | \theta, z_{1:k}, T_{k-1} ) p(X_{k-1} | z_{1:k-1}, y_{1:k-1} ) \, d T_{k-1} \\
& =& \int \left( \int p(T_{k} | \theta, u_{k}, T_{k-1} ) p(u_{k}|z_{1:k}) \, du_{k} \right) p(X_{k-1} | z_{1:k-1}, y_{1:k-1} ) \, d T_{k-1}.
\end{eqnarray*}

\item Analysis step:
\begin{eqnarray*}
p(X_{k} | z_{1:k}, y_{1:k} ) & \propto & p(y_{k} | X_{k}) p(X_{k}| z_{1:k}, y_{1:k-1} ) \\
& = & p(y_{k} | T_{k}) p(X_{k}| z_{1:k}, y_{1:k-1} )\,.
\end{eqnarray*}

\end{enumerate}


 
\subsection{EnMKF for $\theta$ and $T_{k}$}
\label{EnMKF}

Here, the main idea is to consider an ensemble version of the marginalized Kalman filter introduced in \ref{margkalman} to approximate the Bayesian filter \ref{MBF}. We initialize the algorithm by drawing independently $M$ samples from the prior distributions of the parameters and the initial state, $\{ \theta^i \}_{i=1}^{M}$ and $\{ T_0^i \}_{i=1}^{M}$. Then, in the prediction step, we evolve each state vector $T_0^i$ in time given the parameters $\theta^i$ to obtain the state vectors at time $t_1$. The measurements at time $t_1$ are then utilized in the analysis step to update both states and parameters. We iterate the prediction and analysis steps until we use all measurements and reach the final time point $t_N$.

The evolution of the joint vector, $X_k$, is nonlinear because $T_k$ depends on $\theta$ nonlinearly. Therefore, we can not use Kalman filter directly. The ensemble Kalman framework is introduced to handle the nonlinear dependence on $\theta$, and it uses a Gaussian approximation to update the ensemble members in the analysis step. In our EnMKF algorithm, we exploit the conditional linearity of the joint vector to derive its prediction mean and covariance. 

Following the stochastic EnKF scheme \cite{understandEnKF,evensenENKF2}, we assume a trivial evolution of the static parameters and denote the samples from $p(\theta| y_{1:k-1})$ by $\theta^i_{|k-1}$. Given the ensemble size $M$, we compute the predicted state $T^{i}_{k|k-1} = E[ T^i_{k} | \theta_{|k-1}^i, z_{1:k}, y_{1:k-1} ]$ for each $i = 1, \ldots, M$ as
 
\begin{equation}
T^{i}_{k|k-1} = A_{\theta_{|k-1}^{i}} T^{i}_{k-1|k-1} + B_{\theta_{|k-1}^{i}} u_{k|k}.
\label{predstep}
\end{equation}
The prediction covariance matrix, $\mathcal{P}_{k|k-1}$, is usually approximated by the sample covariance matrix of $X_{k|k-1}$,

\begin{equation}
\frac{1}{M-1} \sum_{i=1}^{M} (X^{i}_{k|k-1}-\bar{X}_{k|k-1})(X^{i}_{k|k-1}-\bar{X}_{k|k-1})',
\label{SampleCov}
\end{equation} 
where $\bar{X}_{k|k-1} = \frac{1}{M}\sum_{i=1}^{M} X^{i}_{k|k-1}$. This approximation is needed due to the nonlinearity. Instead, we derive the covariance matrix of $X_{k|k-1}$ using the law of total covariance by conditioning on $\theta_{|k-1}$:
\begin{eqnarray}
\mathcal{P}_{k|k-1} &=& Cov(X_{k}|z_{1:k}, y_{1:k-1}) \nonumber \\
& = &  Cov_{\Theta}(E[X_{k}|\theta_{|k-1}, z_{1:k}, y_{1:k-1}]) + E_{\Theta}[Cov(X_{k}|\theta_{|k-1},  z_{1:k}, y_{1:k-1})] \nonumber \\
& = &  Cov_{\Theta}\left(E \left[ \begin{array}{c} \theta \\ T_k  \end{array} \Big| \theta_{|k-1}, z_{1:k}, y_{1:k-1}\right]\right) + E_{\Theta}\left[Cov\left( \begin{array}{c} \theta \\ T_k  \end{array} \Big| \theta_{|k-1},  z_{1:k}, y_{1:k-1}\right)\right] \nonumber \\
& = &  Cov_{\Theta}\left( \begin{array}{c} \theta_{|k-1} \\ E[T_k |\theta_{|k-1}, z_{1:k}, y_{1:k-1}] \end{array} \right) + E_{\Theta}\left[ \begin{array}{cc} \mathbf{0} & \mathbf{0} \\ \mathbf{0} & Cov(T_k|\theta_{|k-1},  z_{1:k}, y_{1:k-1})  \end{array} \right] \nonumber \\
& \approx &  \frac{1}{M-1}\sum_{i =1 }^{M} \left[ \begin{array}{cc} (\theta_{|k-1}^{i} - \overline{\theta}_{|k-1})(\theta_{|k-1}^{i} - \overline{\theta}_{|k-1})' & (\theta_{|k-1}^{i} - \overline{\theta}_{|k-1})(T^{i}_{k|k-1}- \overline{T}_{k|k-1})' \\ (T^{i}_{k|k-1}- \overline{T}_{k|k-1})(\theta_{|k-1}^{i} - \overline{\theta}_{|k-1})' & (T^{i}_{k|k-1}- \overline{T}_{k|k-1})(T^{i}_{k|k-1}- \overline{T}_{k|k-1})' \end{array} \right] \nonumber \\
& + & \frac{1}{M} \sum_{i=1}^{M} \left[ \begin{array}{cc} \mathbf{0} & \mathbf{0} \\ \mathbf{0} & A_{\theta_{|k-1}^i} P^{i}_{k-1|k-1} A'_{\theta_{|k-1}^i} + B_{\theta_{|k-1}^i} P^{u}_{k|k} B'_{\theta_{|k-1}^i} + W_{k}  \end{array} \right], \label{FullCov}
\end{eqnarray}
where $\overline{\theta}_{|k-1} = \frac{1}{M}\sum_{i=1}^{M}\theta_{|k-1}^{i}$, and $P^{i}_{k-1|k-1}$ must be computed using the Kalman update equations:
\begin{eqnarray*}
&& P^{i}_{k|k} = (I - P^{i}_{k|k-1}H'(HP^{i}_{k|k-1}H + V_k)^{-1} H) P^{i}_{k|k-1}, \\
&& P^{i}_{k|k-1} = A_{\theta_{|k-1}^i} P^{i}_{k-1|k-1} A'_{\theta_{|k-1}^i}.
\end{eqnarray*}

Computing and storing these covariance matrices $P^{i}_{k|k}$ and $P^{i}_{k|k-1}$ for each $i = 1, \ldots, M$ may be burdensome, especially when $\dim(T_k)$ is large. Therefore, we consider the following approximation of the prediction covariance matrix:

\begin{equation}
\label{margjointcov}
\mathcal{P}_{k|k-1} \approx \frac{1}{M-1} \sum_{i=1}^{M}  (X^{i}_{k|k-1}-\bar{X}_{k|k-1})(X^{i}_{k|k-1}-\bar{X}_{k|k-1})' + \frac{1}{M} \sum_{i=1}^{M}  \begin{bmatrix} \mathbf{0} &  \mathbf{0} \\
\mathbf{0}  &  B_{\theta^{i}_{|k-1}}  P^{u}_{k|k} B_{\theta^{i}_{|k-1}}' + W_{k} \end{bmatrix},
\end{equation}
which requires the computation of $B_{\theta^{i}_{|k-1}}  P^{u}_{k|k} B_{\theta^{i}_{|k-1}}'$ for each $i = 1, \ldots, M$ in addition to the standard prediction covariance of EnKF \eqref{SampleCov}. The additional term in \eqref{margjointcov} can be interpreted as inflation with respect to $u_k$ and $w_k$. The full prediction covariance \eqref{FullCov} provides an extra inflation with respect to $T_{k-1}$. However, it is computationally intractable except in very simple cases. The standard practice in EnKF is to consider the sample covariance \eqref{SampleCov} only. Our implementation \eqref{margjointcov} inflates the covariance in a tractable way. In the next section, we introduce another algorithm where we sample $u_k$; therefore, the inflation is not needed.  

After applying the prediction step in \eqref{predstep} and \eqref{margjointcov}, the analysis step for $X_k$ is computed by
\begin{eqnarray}
&& K_k = \mathcal{P}_{k|k-1}\mathcal{H}' \left(\mathcal{H} \mathcal{P}_{k|k-1} \mathcal{H}' + V_{k} \right)^{-1}, \label{kalmangain} \\
&& X^{i}_{k|k} = X^{i}_{k|k-1} + K_{k} \left(y_{k}+ v^{i}_{k} - \mathcal{H} X^{i}_{k|k-1} \right), \label{update}
\end{eqnarray}
where $\mathop{\cal{H}}\limits_{m \times (p+n)} = [ \mathop{\bf{0}}\limits_{m \times p} \quad \mathop{H}\limits_{m \times n} ]$ is the observation operator that maps the augmented vector $X_k$ to the corresponding observation space and $v^{i}_{k} \sim N(0, V_{k})$ is used to perturb the observations. 

A clear advantage of EnMKF is that it runs sequentially on time without the need to restart from the initial state. At a given time $t_k$, the static parameters are updated in the analysis step using measurements at time $t_k$. We summarize the complete algorithm of EnMKF in Algorithm \ref{alg1}. 

\begin{algorithm}[H]
\caption{Ensemble-marginalized Kalman filter (EnMKF)} \label{alg1}
\begin{algorithmic}[1]
\State \textbf{draw} an initial ensemble $\theta^{i} \,,i = 1, \ldots, M$ from the prior distribution $\rho_{\Theta}(\theta)$
\State \textbf{draw} independently an initial ensemble $T_{0}^{i} \,,i = 1, \ldots, M$ from a normal prior distribution $N(T_0,P_0)$
\For{$k= 1$ \textbf{to} number of time observation $N$}  
\State \textbf{run} the Kalman filter for $u_{k}$ (equations \eqref{kalmanpredstep} and \eqref{kalmananalysis}) to obtain $u_{k|k}$ and $P^{u}_{k|k}$ 
\State \textbf{run} the prediction step for $i = 1, \ldots, M$
\[ T^{i}_{k|k-1} = A_{\theta_{|k-1}^{i}} T^{i}_{k-1|k-1} + B_{\theta_{|k-1}^{i}} u_{k|k} \]
\State \textbf{compute} the prediction covariance matrix $\mathcal{P}_{k|k-1}$ using \eqref{margjointcov}
\State \textbf{run} the analysis step for each $X^i, i = 1, \ldots, M$
\begin{eqnarray*}
&& K_k = \mathcal{P}_{k|k-1}\mathcal{H}' \left(\mathcal{H} \mathcal{P}_{k|k-1} \mathcal{H}' + V_{k} \right)^{-1}, \\
&& X^{i}_{k|k} = X^{i}_{k|k-1} + K_{k} \left(y_{k}+ v^{i}_{k} - \mathcal{H} X^{i}_{k|k-1} \right), v^{i}_{k} \sim N(0, V_{k})
\end{eqnarray*}
\EndFor
\end{algorithmic}
\end{algorithm}

In general, ensemble Kalman filters suffer from bias errors, especially with nonlinear systems. A common bias error is due to the Kalman formula that assumes Gaussianity. Nevertheless, the incorporation of the boundary temperature uncertainties will reduce the total bias error in the estimated parameters \cite{sawlan2}. We show in subsection \ref{biaserror} that EnMKF admits smaller bias errors than the following EnKF algorithm.

\subsection{Ensemble Kalman filter for $\theta$ and $T_{k}$}
\label{EnKF}

An alternative approach to solve our problem \eqref{system1}--\eqref{system2}, is to apply an adequately modified EnKF to the joint vector $X_k = \left[\, \theta \,\,\,\,\, T_k \, \right]'$. In this modified EnKF, the control vector $u_k$ is sampled from its distribution, which was determined during the prediction step. Thus, we sample $u^{i}_{k}$ from $N(u_{k|k}, P^{u}_{k|k})$ for each $i = 1, \ldots, M$. Then, the prediction step differs from \eqref{predstep} and is given by

\begin{equation*}
T^{i}_{k|k-1} = A_{\theta_{|k-1}^{i}} T^{i}_{k-1|k-1} + B_{\theta_{|k-1}^{i}} u^{i}_{k}\,.
\end{equation*}
In this case, the prediction covariance matrix is approximated by

\begin{equation}
\label{EnKFcov}
\frac{1}{M-1} \sum_{i=1}^{M} (X^{i}_{k|k-1}-\bar{X}_{k|k-1})(X^{i}_{k|k-1}-\bar{X}_{k|k-1})' + \begin{bmatrix} \mathbf{0} &  \mathbf{0} \\
\mathbf{0}  &  W_{k} \end{bmatrix}.
\end{equation}
The analysis step is the same as the one that we developed for EnMKF, which is summarized by equations \eqref{kalmangain} and \eqref{update}.

Both EnMKF and EnKF are Monte Carlo implementations of Bayesian filtering. The main difference appears in the prediction step, especially in the integration with respect to $u_k$. In EnMKF, we use the fact that $u_k|z_{1:k}$ and $T_{k} | \theta, u_{k}, T_{k-1}$ have Gaussian distributions and, therefore, the resulting distribution is Gaussian with mean \eqref{margmean} and covariance \eqref{margcov}. In EnKF \ref{EnKF}, the same integral is approximated with the Monte Carlo method by sampling from $p(u_k|z_{1:k})$.   

The modified EnKF approach avoids the computation of the matrices $B_{\theta_{|k-1}^i}  P^{u}_{k|k} B_{\theta_{|k-1}^i}'$ that are needed in formula \eqref{margjointcov} of EnMKF. However, the additional term in EnMKF can be considered inflation induced from the law of total covariance by conditioning on the unknown parameters $\theta$. Similarly, the joint distribution of the state and parameters can be split into the distribution of the conditional state and the marginal distribution of the parameters \cite{BadaptEnKF}. In our case, the linear evolution of the state and control vectors simplifies the conditional distribution to the conditional mean \eqref{margmean} and the conditional covariance \eqref{margcov}.  


\section{Real-world application: Heat transfer across a solid brick wall}
\label{application}

In this section, we show how to formulate and implement the EnMKF method to deal with the state and parameter estimation problem in a real-world heat-transfer application, which is summarized in subsection \ref{sub6-1}. Next, we compare the results obtained with the EnMKF Algorithm \ref{alg1} to those obtained with the modified EnKF method from Section \ref{EnKF} using experimental and synthetic data. We also propose stopping criteria that indicate when the ensemble parameters are stationary with adequate non-zero variance.

We consider the problem of describing the thermal performance of a solid wall given in-situ measurements. This can be posed as a state and parameter estimation problem constrained by the heat equation. For simplicity, we consider the case of a single-layer wall with homogeneous density, thermal conductivity, and specific heat capacitance. Therefore, we characterize the thermophysical properties of the wall in terms of its heat capacity per unit area  $(J/m^2 K)$ and thermal resistance $(m^2 K/W)$. These are denoted by $\rho C$ and $R$, respectively. Under the standard assumption of uni-directional heat flux across the wall's thickness, the forward model is described by the internal temperature profile of the wall, denoted by $T(x,t)$. This profile is the solution to the 1D heat equation \cite{yunus2011heat}:

\begin{equation}
\begin{cases}
\frac{\rho C}{L} \frac{\partial T}{\partial t} =  \frac{\partial}{\partial x} \left( \frac{L}{R} \, \frac{\partial T}{\partial x} \right),\,  x \in (0, L),\,  t \in [0,t_{N}] \\
T(0,t) = T_{int}(t),\,  t \in [0,t_{N}] \\
T(L,t) = T_{ext}(t),\,  t \in [0,t_{N}] \\
T(x,0) = T_0(x),\, x \in (0, L), 
\end{cases}
\label{heateq0}
\end{equation}
where $L$ is the wall thickness, $T_{int}$ and $T_{ext}$ are the internal and external wall surface temperatures, respectively, and $T_{0}(x)$, is the initial temperature of the wall. The goal is to estimate the thermophysical properties $\theta = ( R, \rho C)$ given boundary temperature measurements $T_{int}$ and $T_{ext}$ as well as measurements of the internal and external boundary heat flux defined via:

\begin{equation}
\begin{cases}
F_{int}(t) = & -\frac{L}{R} \frac{\partial T}{\partial x}|_{x=0}, \\
F_{ext}(t) = & -\frac{L}{R} \frac{\partial T}{\partial x}|_{x=L}.
\end{cases}
\label{heateq01}
\end{equation}
An experimental setting consistent with (\ref{heateq0})-(\ref{heateq01}) can be designed by placing temperature sensors and heat flux meters on the internal and external surfaces of the wall \cite{sawlan2,REnKA}. These quantities are then monitored for a period of time which often consist of at least two weeks. In order to reduce the effect of solar radiation (not considered in (\ref{heateq0})), measurement campaigns are usually conducted on north-facing walls during the winter \cite{ISO9869:2014}. For simplicity, we also assume that the initial condition, $T_{0}(x)$, is well-approximated by the piecewise linear function

\begin{equation}\label{model_as}
\begin{cases} 
      T_{int}(0) + 2 \frac{\tau_{0} - T_{int}(0)}{L} x & \textrm{ if $0 < x \leq \frac{L}{2}$} \\
      \tau_0 + 2\frac{T_{ext}(0) - \tau_{0}}{L} (x - \frac{L}{2}) & \textrm{ if $\frac{L}{2} < x < L$} \, ,
   \end{cases}
\end{equation}
where we assume $\tau_0 = 16.1^{\circ} C$, according to the estimate we obtained in our previous study \cite{sawlan2}. 

To formulate the discrete state observation system, we consider the following uniform space-time discretization:

\[ x_0 = 0, x_1 =  \Delta x, \ldots, x_i = i \Delta x, \ldots, x_n = n \Delta x = L \, , \] 
\[ t_0 = 0, t_1 =  \Delta t, \ldots, t_j = j \Delta t, \ldots, t_N = N\Delta t \, , \] 
and denote by ${\hat{T}_{k}} = \left( T(\Delta x, k \Delta t), \ldots, T((n-1)\Delta x, k \Delta t) \right), T_{int, k} = T_{int} (k\Delta t)$ and $T_{ext, k} = T_{ext} (k\Delta t)$ the inner, internal and external surface temperatures at time $t_k$, respectively. The inner temperature vector $\hat{T}_{k}$ can be written as a linear function of $T_{int,k}$ and $T_{ext,k}$ as follows \cite{sawlan2}:

\begin{equation*}
\hat{T}_{k} = \tilde{A_{\theta}} \hat{T}_{k-1} + \tilde{B}_{\theta, int} T_{int,k} + \tilde{B}_{\theta, ext} T_{ext,k},
\end{equation*}
where the matrices $\tilde{A}_{\theta}, \tilde{B}_{\theta, int}$, and $\tilde{B}_{\theta, ext}$ are specified exactly (see \cite{sawlan2}).

We let $T_k = [ \begin{array}{ccc} T_{int,k} & \hat{T}_{k}' & T_{ext,k} \end{array} ]'$ be the temperature vector on the closed interval $[0,L]$ and

\[ A_{\theta} = \begin{bmatrix} 0 \\ {\tilde{A}}_{\theta} \\ 0 \end{bmatrix}, \, B_{\theta, int} = \begin{bmatrix} 1 \\ \tilde{B}_{\theta, int} \\ 0 \end{bmatrix}, \, B_{\theta, ext} = \begin{bmatrix} 0 \\ \tilde{B}_{\theta, ext} \\ 1 \end{bmatrix}. \]
Then, the time evolution equation of the state vector $T_k$ is given by
\begin{equation} \label{state}
T_{k} = A_{\theta} T_{k-1} + B_{\theta, int} T_{int,k} + B_{\theta, ext} T_{ext,k}.
\end{equation}

The heat-flux vector $F_k =  \left[ \begin{array}{cc} F_{int,k} & F_{ext,k} \end{array} \right]'$ is approximated by the first difference

\[ F_{int,k} = \frac{1}{R} \, \frac{1}{2 \Delta x} ( 3 T_{int,k} - 4 T_{1,k} + T_{2,k}) \]
\[ F_{ext,k} = \frac{1}{R} \, \frac{1}{2 \Delta x} ( 3 T_{ext,k} - 4 T_{n-1,k} + T_{n-2,k}). \]
Therefore, the observation matrix operator takes the form

\[ H = \frac{1}{2 \Delta x} \begin{bmatrix} 3 & -4 & 1 & 0 & \dots & 0 \\
 0 & \dots & 0 & 1 & -4 & 3 \end{bmatrix} \]
 and the observation equation is

 \begin{equation}\label{observation}
 y_k = \frac{1}{R} H T_k + v_k,
 \end{equation}
 where $v_k \sim N(0, V_k)$ is the measurements noise and $\mathcal{H} = [ \begin{array}{cc} \bold{0} & H \end{array}]$ is the observation operator that maps the augmented vector $X_k$ to the observation space. 

We define the observation matrix operator without $R$ to avoid dependence of the Kalman gain on this unknown parameter. Instead, we keep the Kalman gain fixed in each iteration of the algorithm and modify the so-called innovation to be $R y_k - H T_k$. We tailor the EnMKF in Algorithm \ref{alg1} to the state observation system \eqref{state}-\eqref{observation} in Algorithm \ref{alg2}, by augmenting the state vector with $\log(R)$ and $\log(\rho C)$ to ensure that the physical parameters are positive.

The heat conduction problem described in \eqref{heateq0} for a single-layer wall model is relevant for the experimental setting described in the following subsection. However, it is important to emphasize that the EnMKF algorithm introduced in subsection \ref{EnMKF} can be used to infer thermophysical properties of a much wider class of walls. Multilayer walls, for example, can be described by suitable modifications to \eqref{heateq0} that incorporate the thermophysical properties on each layer. Each of these properties can then be included in the unknown vector of parameters $\theta$ that is inferred within the EnMKF. 

\begin{algorithm}[h!]
\caption{EnMKF algorithm to estimate $R$, $\rho C$, $F_{int,k}$ and $F_{ext,k}$} \label{alg2}
\begin{algorithmic}[1]
\State \textbf{draw} an initial ensemble $\theta^{i} =  ( R^{i}, \rho C^{i})\,,i = 1, \ldots, M$ from a prior distribution $\rho_{\Theta}(\theta)$
\State \textbf{approximate} the initial condition $T_{0}$ using \eqref{model_as}
\State \textbf{draw} an initial ensemble $T_{0}^{i} \,,i = 1, \ldots, M$ from the normal prior distribution $N(T_0, P_0)$
\State \textbf{define} the augmented ensemble $X^{i} =  \left[ \begin{array}{ccc} \log(R^{i}) & \log(\rho C^{i}) &  {T^{i}}' \end{array} \right] '\,, i = 1, \ldots, M$
\For{$k= 1$ \textbf{to} number of time observation $N$}  
\State \textbf{run} the Kalman filter for $T_{int,k}$ and $T_{ext,k}$ to obtain $T_{int,k|k}, T_{ext,k|k}, P_{int,k|k}$ and $P_{ext,k|k}$
\State \textbf{run} the prediction step for $i = 1, \ldots, M$
\[ T^i_{k|k-1} = A_{\theta^i_{|k-1}} T^i_{k-1|k-1} + B_{\theta^i_{|k-1}, int} T_{int,k|k} + B_{\theta^i_{|k-1}, ext} T_{ext,k|k} \]
\State \textbf{compute} the prediction covariance matrix
\[ \mathcal{P}_{k|k-1} \approx \frac{1}{M-1} \sum_{i=1}^{M}  (X^{i}_{k|k-1}-\bar{X}_{k|k-1})(X^{i}_{k|k-1}-\bar{X}_{k|k-1})' \]
\[ + \frac{1}{M} \sum_{i=1}^{M}  \begin{bmatrix} \mathbf{0} &  \mathbf{0} \\
\mathbf{0}  &  B_{\theta^{i}_{|k-1},int}  P_{int,k|k} B_{\theta^{i}_{|k-1},int}' + B_{\theta^{i}_{|k-1},ext}  P_{ext,k|k} B_{\theta^{i}_{|k-1},ext}' \end{bmatrix} \]
\State \textbf{run} the analysis step for each $X^i, i = 1, \ldots, M$
\[  K_k = \mathcal{P}_{k|k-1}\mathcal{H}' \left(\mathcal{H} \mathcal{P}_{k|k-1} \mathcal{H}' + V_{k} \right)^{-1} \]
\[ X^{i}_{k|k} = X^{i}_{k|k-1} + K_{k} \left(R^{i}(y_{k}+ v^{i}_{k}) - \mathcal{H} X^{i}_{k|k-1} \right) \,, v^{i}_{k} \sim N(0, V_{k}) \]
\State \textbf{estimate} the heat flux and its sample mean and sample covariance
\[ F^{i}_{k} = \frac{1}{R^i_{|k}} H T^i_{k|k} \]
\[ \widehat{F}_{k} = \frac{1}{M} \sum_{i=1}^{M} F^{i}_{k} \]
\[ {\widehat{\Sigma}}(F_k) = \frac{1}{M-1} \sum_{i=1}^{M} (F^{i}_{k} - \widehat{F}_{k})'(F^{i}_{k} - \widehat{F}_{k}) \]
\EndFor
\end{algorithmic}
\end{algorithm}

\subsection{Experimental data set}
\label{sub6-1}
We use real data from an experiment conducted inside an environmental chamber in the Energy Technologies Building at Nottingham University Innovation Park. The chamber was divided into two rooms by a $215-$mm thick partition wall. The aim of the experiment was to estimate the thermal properties of a brick section of the partition wall. The temperature in both rooms was controlled to resemble internal conditions in Room 1 and external conditions in Room 2. The wall boundary temperature and heat-flux measurements were recorded every minute by sensors placed on the surface of the brick wall \cite{sawlan2}.

\begin{figure}[h!]
\centering
\includegraphics[width=0.8\textwidth]{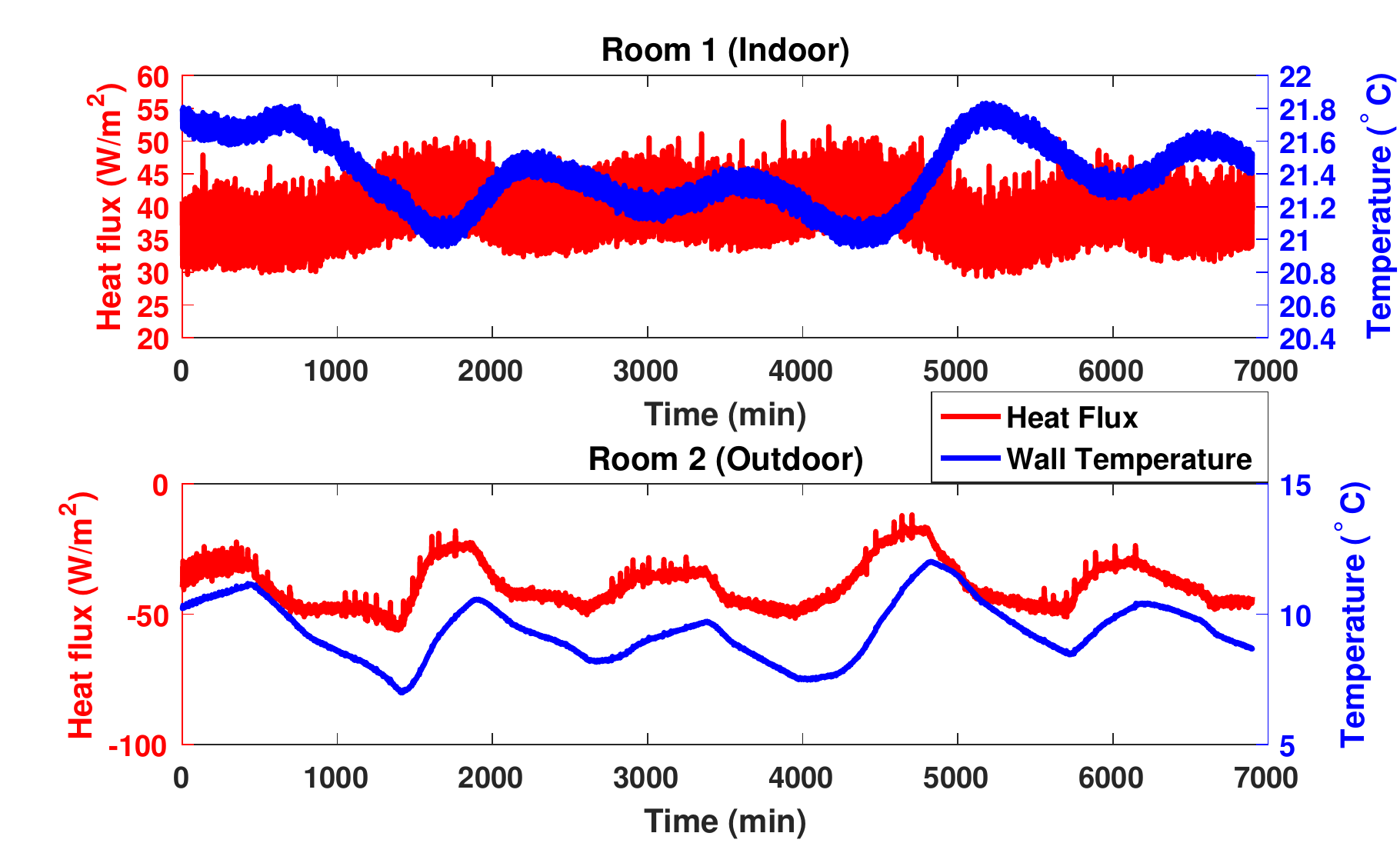}
\caption{Raw temperature and heat-flux measurements. Temperature in Room 2 imitates outdoor weather conditions.}
\label{fig1}
\end{figure}

Figure \ref{fig1} shows the time series of the temperature and heat flux measurements corresponding to the two sides of the brick wall, identified by internal and external measurements. The measurements contain noise that we can analyze, in a non-sequential framework, by using a smoothing spline method to fit a curve to each time series. Then, we approximate the noise by the difference between the measurements and the smooth curve values. The noise variance is estimated to be around $0.01$ for the temperature measurements. The noise variances for the internal and external heat flux measurements are estimated to be $20$ and $5$, respectively. Therefore, the covariance matrix $V_k$ is approximated by 

\[ V_k = \begin{bmatrix} 20 & 0 \\ 0 & 5 \end{bmatrix}\,.\]

Alternatively, we can approximate $V_k$ by using prior knowledge on the accuracy of the measuring devices in different situations or by assuming a prior model with unknown hyper-parameters and estimate them sequentially along with the quantities of interest. Further statistical analyses and details about these experimental data are provided in \cite{sawlan2}.

\subsection{Kalman filtering for $u_{k}$}
\label{results1}

\begin{figure}[h!]
\centering
\begin{minipage}{.48\textwidth}
  \centering
  \includegraphics[width=1.1\textwidth]{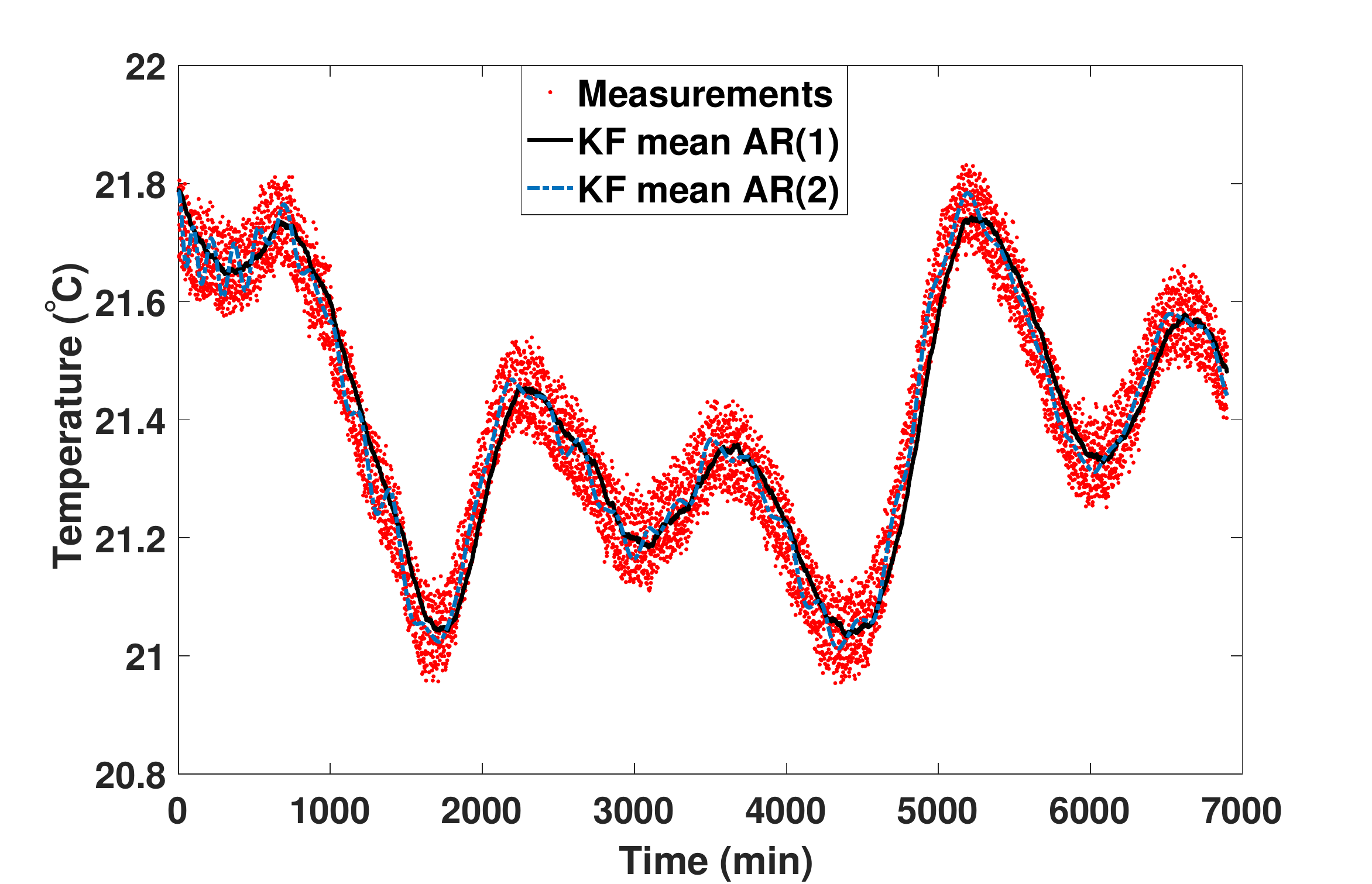}
  \captionof{figure}{Internal temperature measurements (red dots) and Kalman filter mean with AR(1) and AR(2).}
  \label{KFint}
\end{minipage}
~~~~
\begin{minipage}{.48\textwidth}
  \centering
  \includegraphics[width=1.1\textwidth]{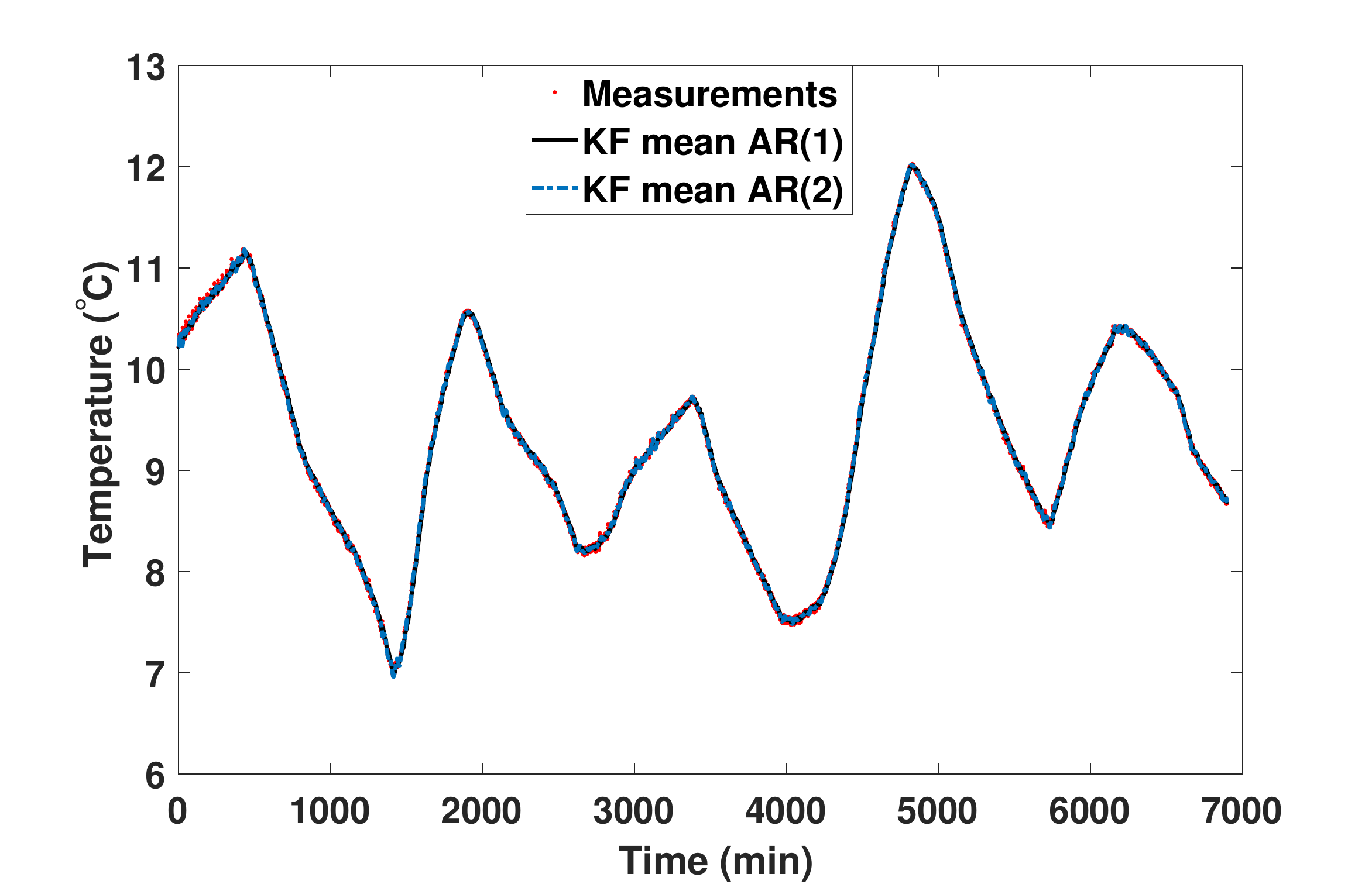}
  \captionof{figure}{External temperature measurements (red dots) and Kalman filter mean with AR(1) and AR(2).}
   \label{KFext}
\end{minipage}
\end{figure}

The first step in implementing EnMKF is to run a Kalman filter for the control vector $u_k$ in \eqref{system2}, which in this case, consists of the boundary conditions $T_{int,k}$ and $T_{ext,k}$. The forward-evolution operator for the boundary conditions is unknown but noisy measurements are available. We consider the two autoregressive models $AR(1)$ and $AR(2)$ from subsection \ref{KF_sec} and use them with the measurements to apply the Kalman filter. Figures \ref{KFint} and \ref{KFext} show the results from the Kalman filter with each model and the real temperature measurements. In the experiments with EnMKF and EnKF presented below, we use only the results from the Kalman filter with $AR(1)$.

\begin{figure}[h!]
\centering
\includegraphics[width=0.65\textwidth]{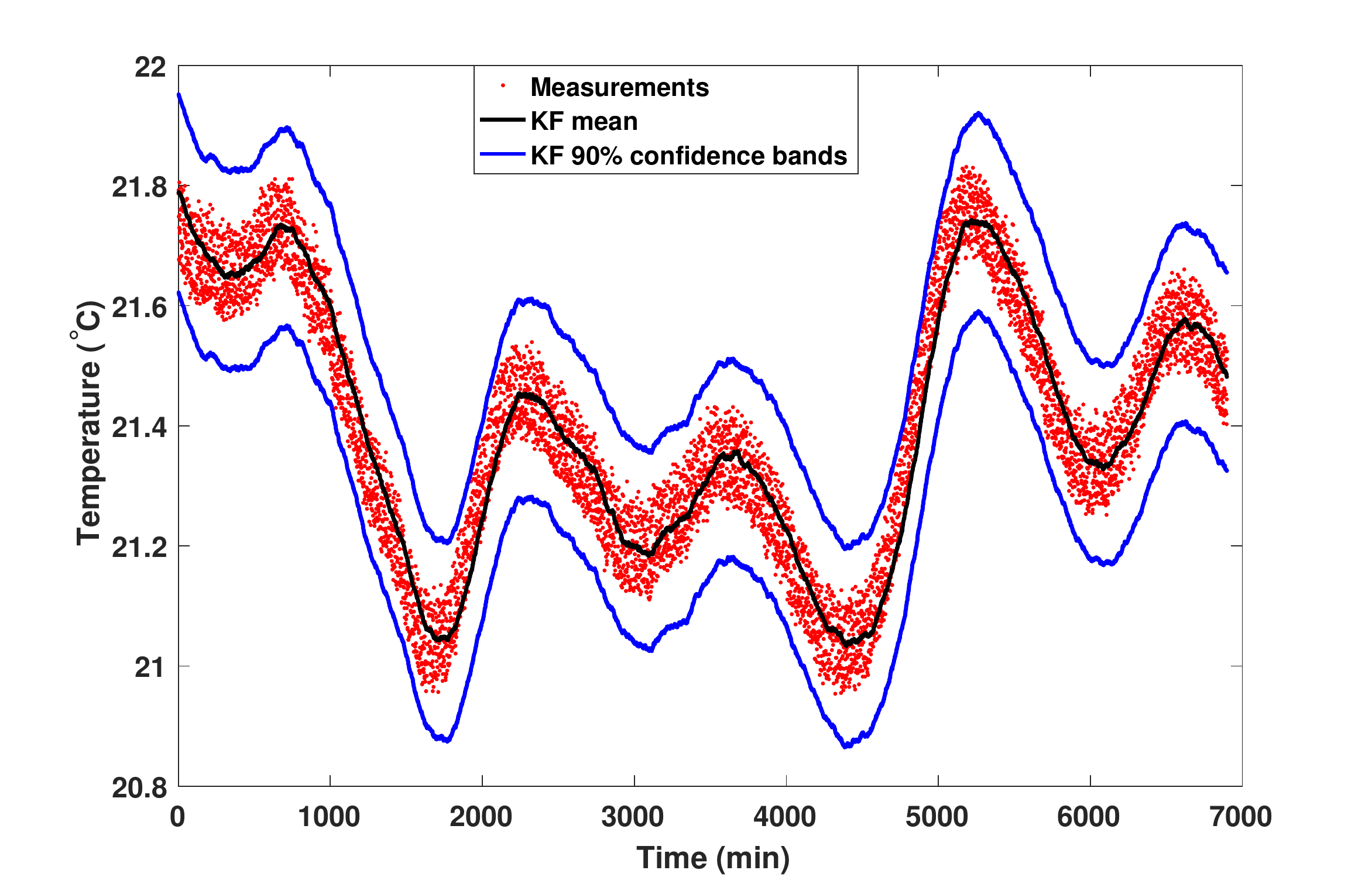}
\caption{Internal temperature measurements (red dots) and Kalman filter mean and $90\%$ confidence bands with AR(1).}
\label{KFintconf}
\end{figure}

We also estimate the variance for the boundary conditions $T_{int,k}$ and $T_{ext,k}$. For example, Figure \ref{KFintconf} shows the estimated mean and confidence bands for the internal temperature measurements using Kalman filter with AR(1). The estimated variances capture the variability of the data and therefore we can marginalize the uncertainty of the boundary conditions accurately.

\subsection{EnMKF results}

We run EnMKF Algorithm \ref{alg2} with $M = 100$ and using the complete experimental data set, $N = 6,900$. We initially sample $R^i$ and $\rho C^i$ from the uniform priors, $U(0.17, 0.36)$ and $U(234000, 431000)$, and sample $T^i_0$ from the normal prior distribution, $N(T_0, 0.01 \mathcal{I}_n)$. 

\begin{figure}[h!]
\centering
\includegraphics[width=1.0\textwidth]{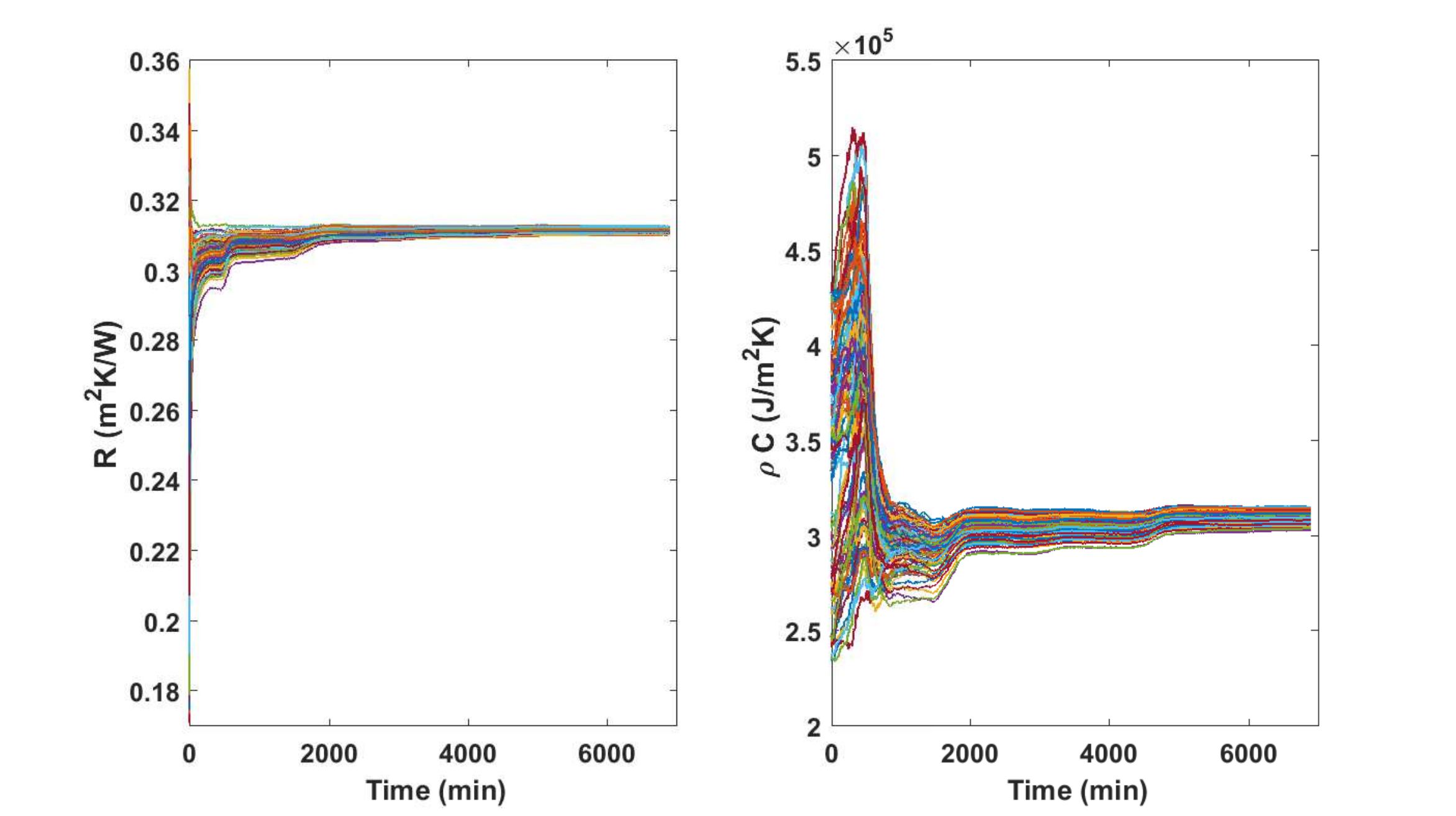}
\caption{Convergence of the thermal resistance $R$ (left) and heat capacity per unit area $\rho C$ (right) ensemble with respect to time using EnMKF with ensemble size $M = 100$.}
\label{realcon}
\end{figure}

Figure \ref{realcon} shows the convergence of the ensemble parameters $R$ and $\rho C$ with respect to time as more observations are incorporated. The thermal-resistance mean converges to $0.31 \, m^2K/W$ and the mean of the heat capacity per unit area converges to $3.11 \times 10^5 \, J/m^2K$. As we increase the ensemble size, we obtain results that are more accurate and consistent with the MCMC results in \cite{sawlan2}. We further analyze the convergence of EnMKF with respect to the ensemble size $M$ and compare it with the convergence of the modified EnKF in subsection \ref{compare}.

\begin{figure}[h!]
\centering
\begin{minipage}{.48\textwidth}
  \centering
  \includegraphics[width=1.1\textwidth]{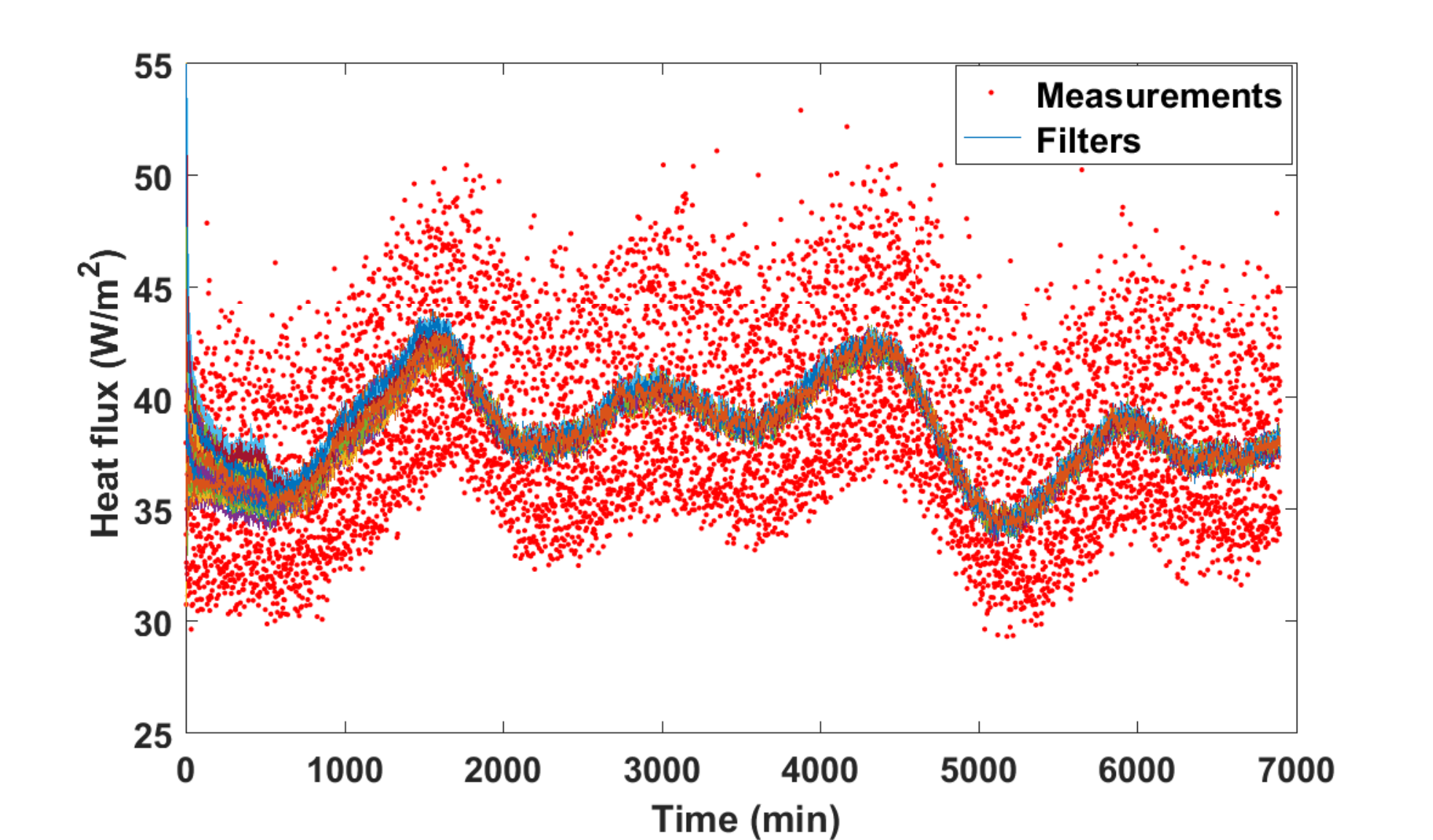}
  \captionof{figure}{Estimated heat flux ensemble (solid lines) using EnMKF with ensemble size $M = 100$ compared with real raw measurements (red dots) in Room 1.}
  \label{HFint}
\end{minipage}
~~~~
\begin{minipage}{.48\textwidth}
  \centering
  \includegraphics[width=1.1\textwidth]{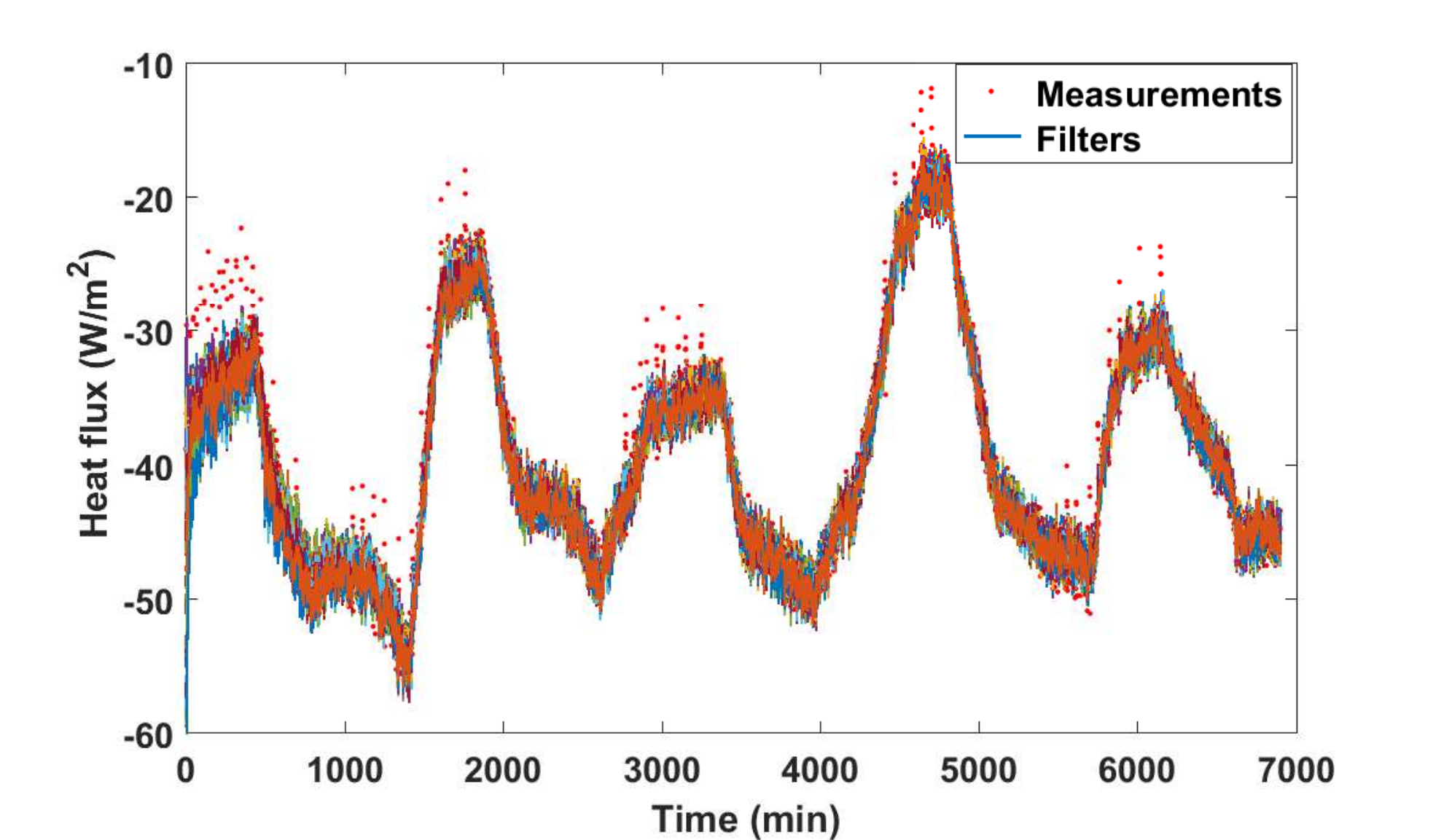}
  \captionof{figure}{Estimated heat flux ensemble (solid lines) using EnMKF with ensemble size $M = 100$ compared with real raw measurements (red dots) in Room 2.}
   \label{HFext}
\end{minipage}
\end{figure}

We also compute the heat flux from the estimated temperature and parameters by EnMKF. Figures \ref{HFint} and \ref{HFext} show that the estimated heat fluxes match the real measurements suitably. The corresponding variances for the estimated heat fluxes converge to be smaller than $1$. We recall that the heat flux measurements are very noisy and it would not be optimal to recover the measurements variance through the variance of the estimates. Figures \ref{HFint} and \ref{HFext} indicate that our estimates are unbiased and therefore smaller variances are preferable.


\subsection{EnKF results}

\begin{figure}[h!]
\centering
\includegraphics[width=1.0\textwidth]{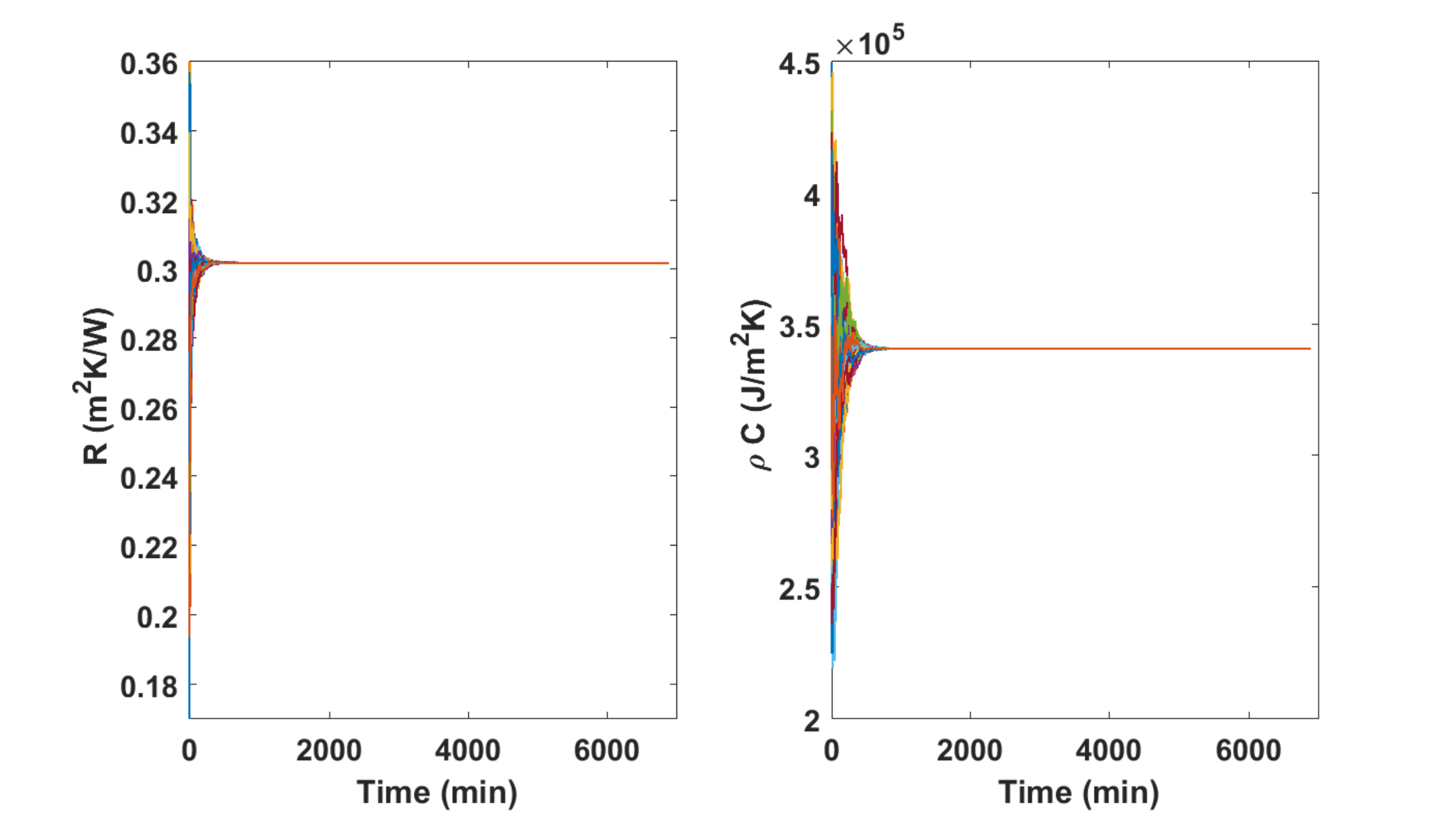}
\caption{Propogation of the thermal resistance $R$ (left) and heat capacity $\rho C$ (right) ensemble with respect to time using EnKF with ensemble size $M = 100$.}
\label{realconbias}
\end{figure}

\begin{figure}[h!]
\centering
\begin{minipage}{.48\textwidth}
  \centering
  \includegraphics[width=1.1\textwidth]{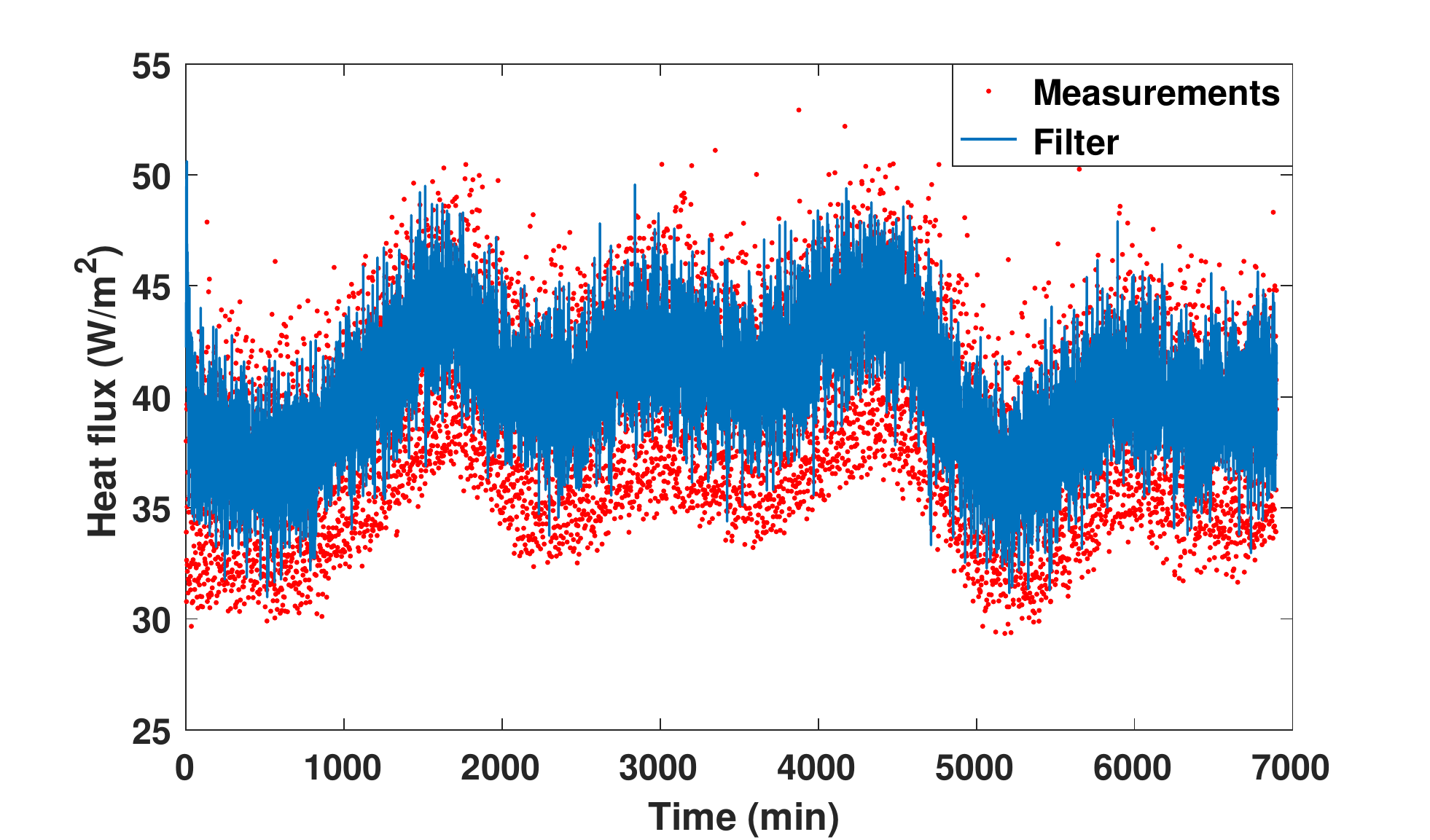}
  \captionof{figure}{Estimated heat-flux mean using EnKF with ensemble size $M = 100$ compared with real raw data measurements in Room 1.}
  \label{HFintb}
\end{minipage}
~~~~
\begin{minipage}{.48\textwidth}
  \centering
  \includegraphics[width=1.1\textwidth]{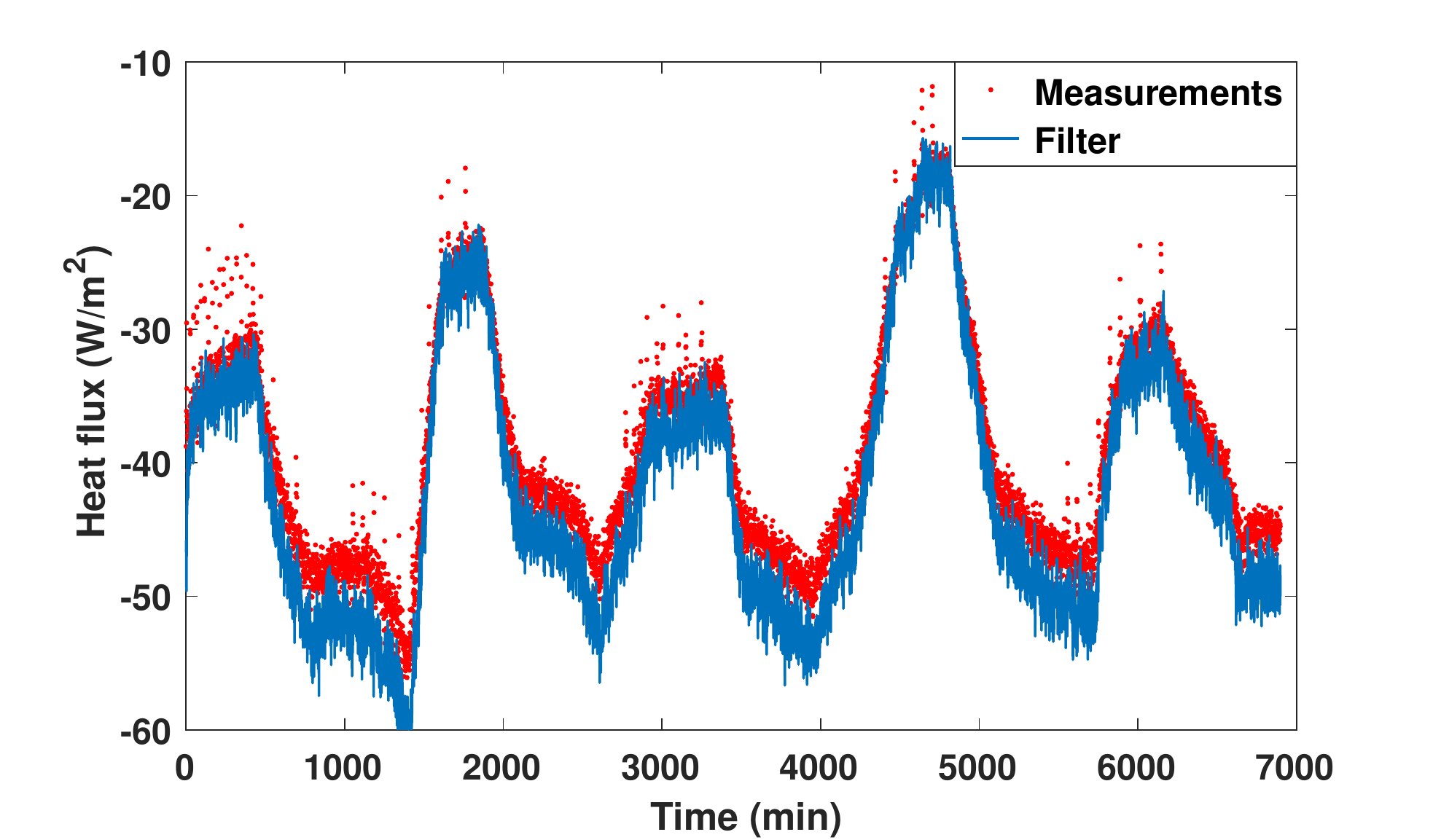}
  \captionof{figure}{Estimated heat-flux mean using EnKF with ensemble size $M = 100$ compared with real raw data measurements in Room 2.}
   \label{HFextb}
\end{minipage}
\end{figure}

We apply the modified EnKF approach introduced in Section \ref{EnKF}, with ensemble size $M = 100$, to infer the parameters $R$ and $\rho C$ as in the previous example. Figure \ref{realconbias} shows that the ensemble of $R$ and $\rho C$ collapse at time $t = 1734$ minutes. After this time, the parameters remain fixed at the biased estimates $R = 0.3\, m^2K/W$ and $\rho C =  3.42 \times 10^5\, J/m^2K$. Thus, the estimated heat fluxes do not match the measurements. In Figures \ref{HFintb} and \ref{HFextb}, we show only the sample mean of the heat-flux ensemble because the variances of the internal and external heat fluxes are approximately $390$ and $130$, respectively. These values are much larger than the estimated variances produced by the EnMKF approach.   

\begin{figure}[h!]
\centering
\includegraphics[width=1.0\textwidth]{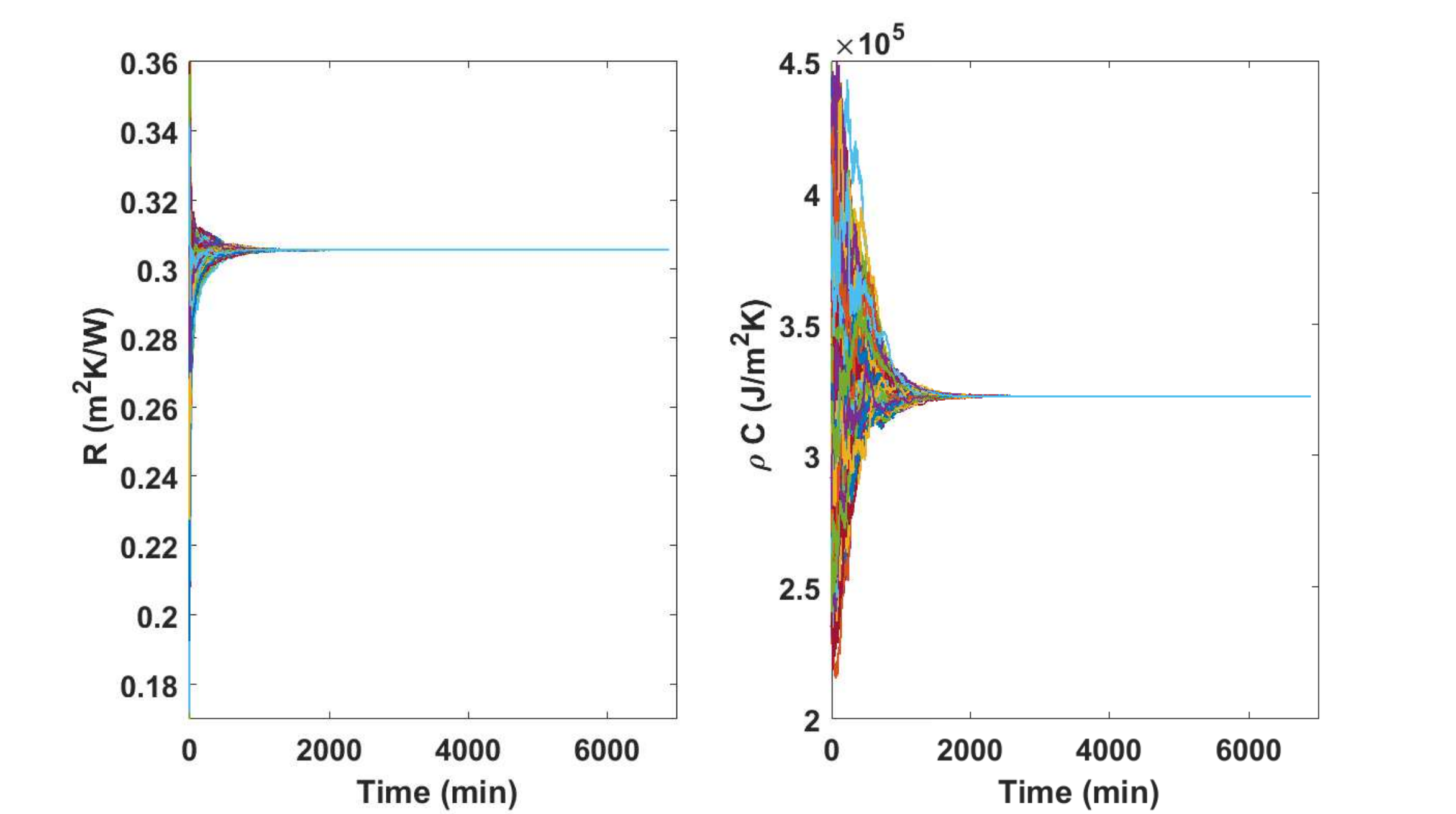}
\caption{Propagation of the thermal resistance $R$ (left) and heat capacity $\rho C$ (right) ensemble with respect to time using EnKF with ensemble size $M = 300$.}
\label{realconbias3}
\end{figure}

\begin{figure}[h!]
\centering
\begin{minipage}{.48\textwidth}
  \centering
  \includegraphics[width=1.1\textwidth]{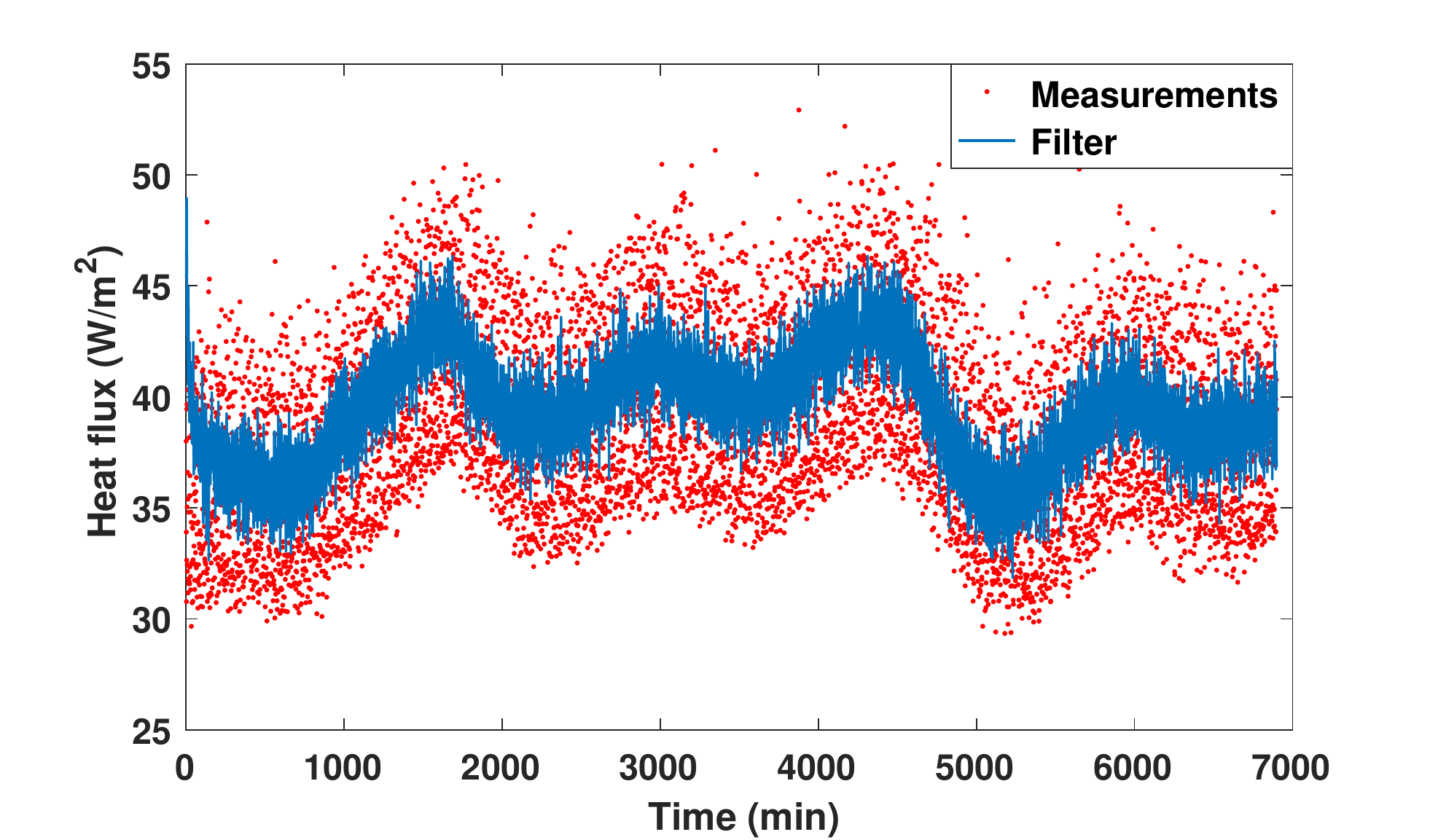}
  \captionof{figure}{Estimated heat-flux mean using EnKF with ensemble size $M = 300$ compared with real raw data measurements in Room 1.}
  \label{HFintb3}
\end{minipage}
~~~~
\begin{minipage}{.48\textwidth}
  \centering
  \includegraphics[width=1.1\textwidth]{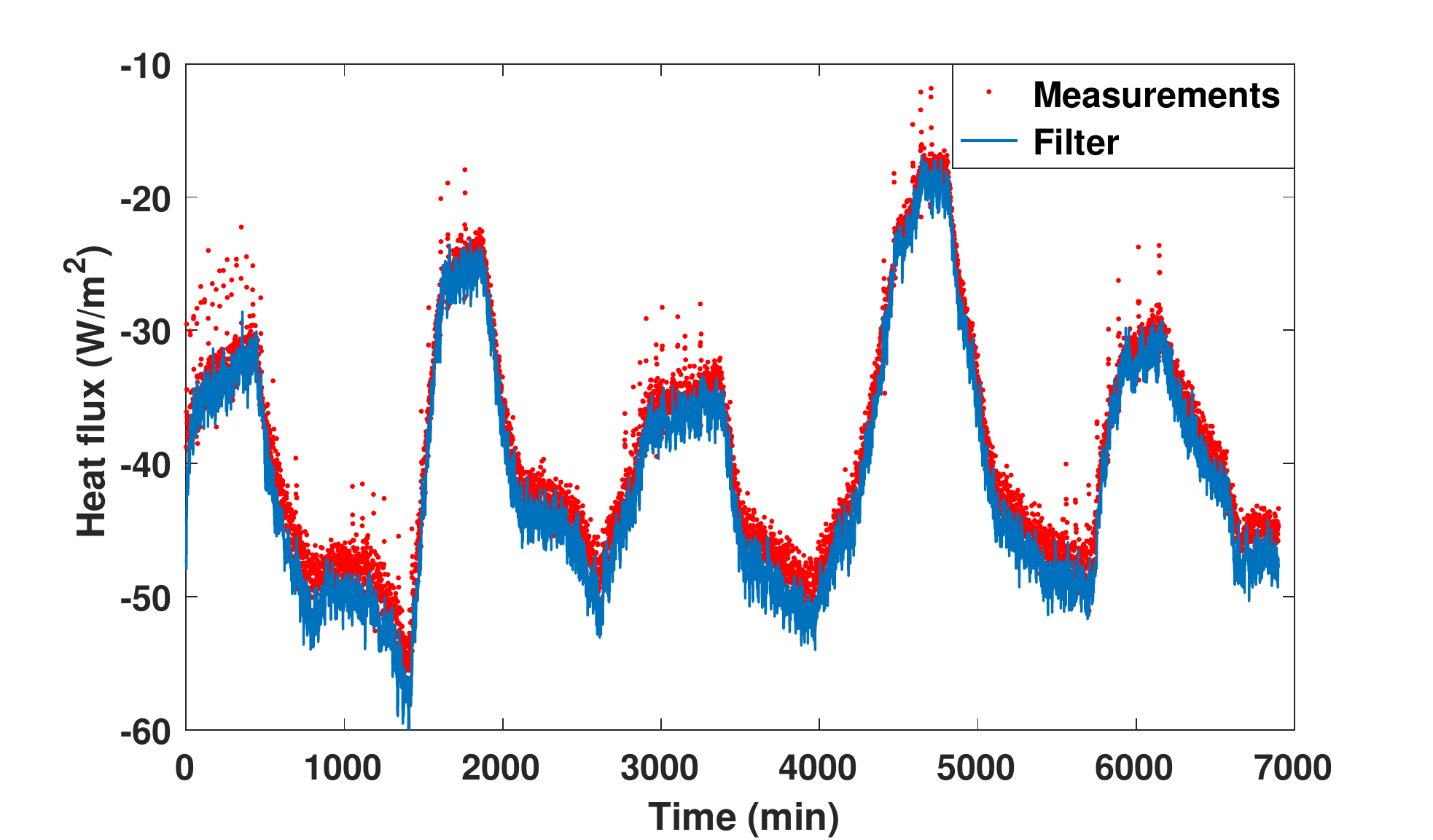}
  \captionof{figure}{Estimated heat-flux mean using EnKF with ensemble size $M = 300$ compared with real raw data measurements in Room 2.}
   \label{HFextb3}
\end{minipage}
\end{figure}

We can gradually increase the ensemble size $M$ to mitigate the ensemble collapse and reduce the bias error in the estimated parameters. Figures \ref{realconbias3}, \ref{HFintb3} and \ref{HFextb3} show the estimated parameters and heat-flux mean when $M = 300$. Although the bias error of the estimated parameters is reduced, the corresponding variances still vanish as time increases, which is not a realistic feature. While the internal and external heat fluxes variances are reduced to $375$ and $124$, respectively, they are still very large values when compared with the estimated variances produced by EnMKF.

\subsection{Comparing EnMKF and EnKF}
\label{compare}
\begin{figure}[h!]
\centering
\includegraphics[scale=0.6]{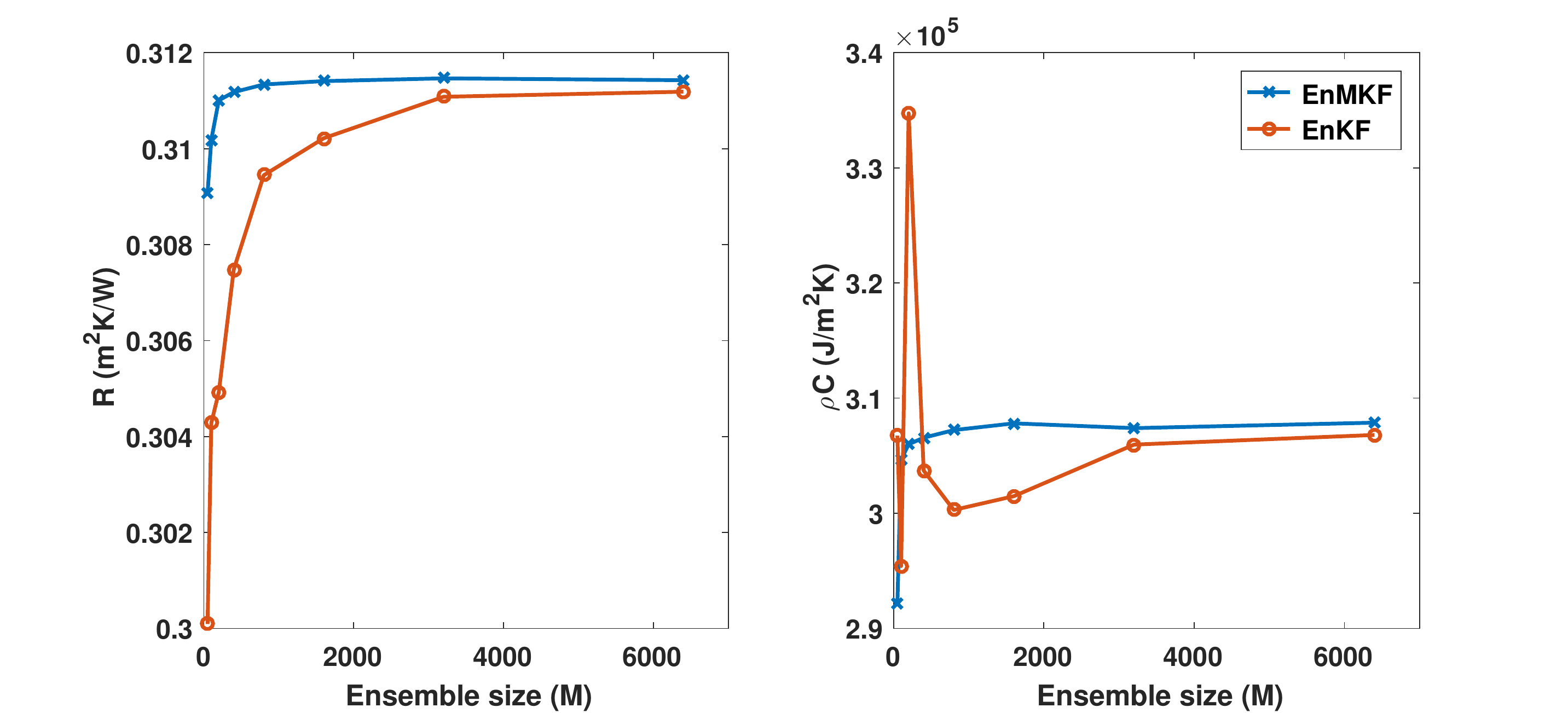}
\caption{Convergence of the ensemble mean of the estimated parameters $R$ (left) and $\rho C$ (right) at time $t' = 3000$ with respect to the ensemble size $M$.}
\label{meanConv}
\end{figure}

\begin{figure}[h!]
\centering
\includegraphics[scale=0.6]{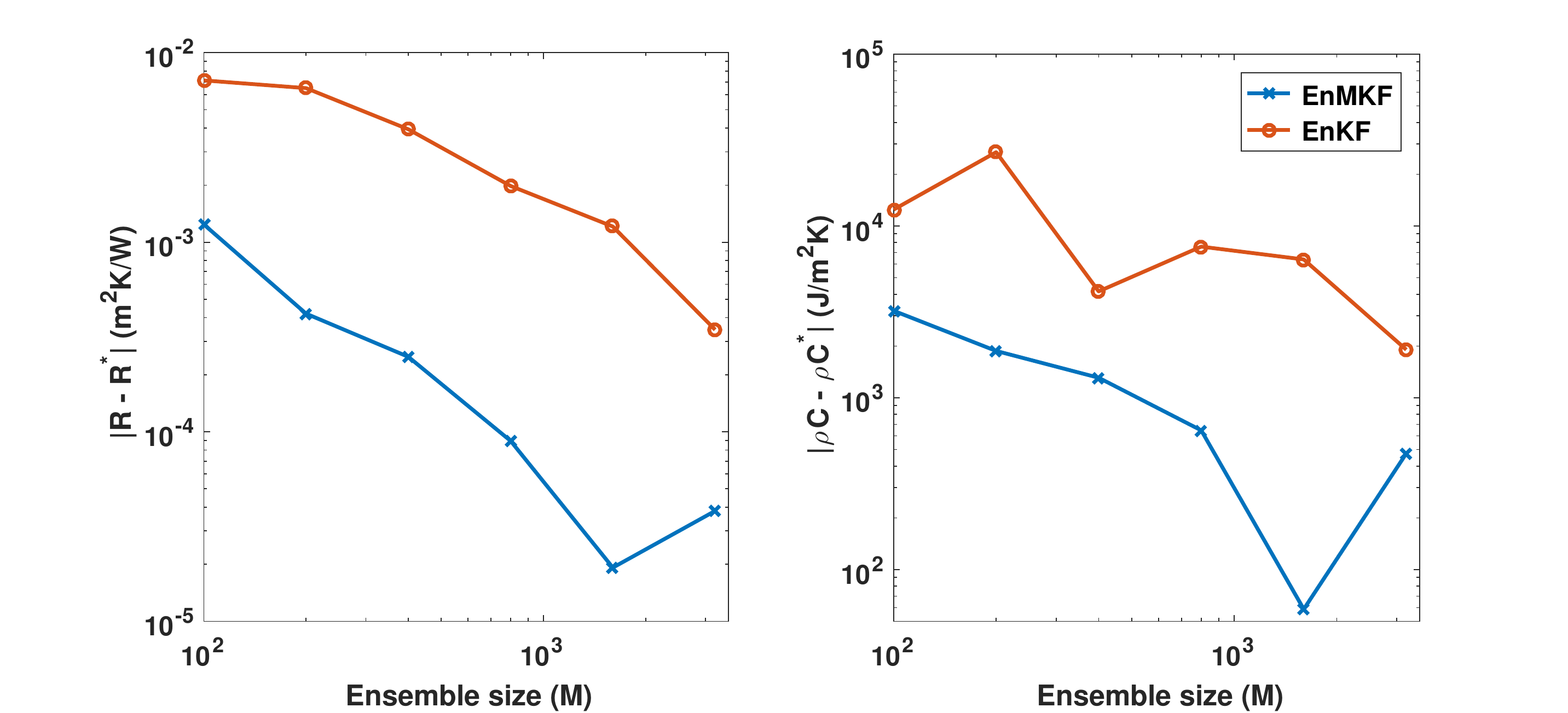}
\caption{Convergence of the ensemble mean error of the estimated parameters $R$ (left) and $\rho C$ (right) at time $t' = 3000$ with respect to the ensemble size $M$.}
\label{meanErrorConv}
\end{figure}

\begin{figure}[h!]
\centering
\includegraphics[scale=0.6]{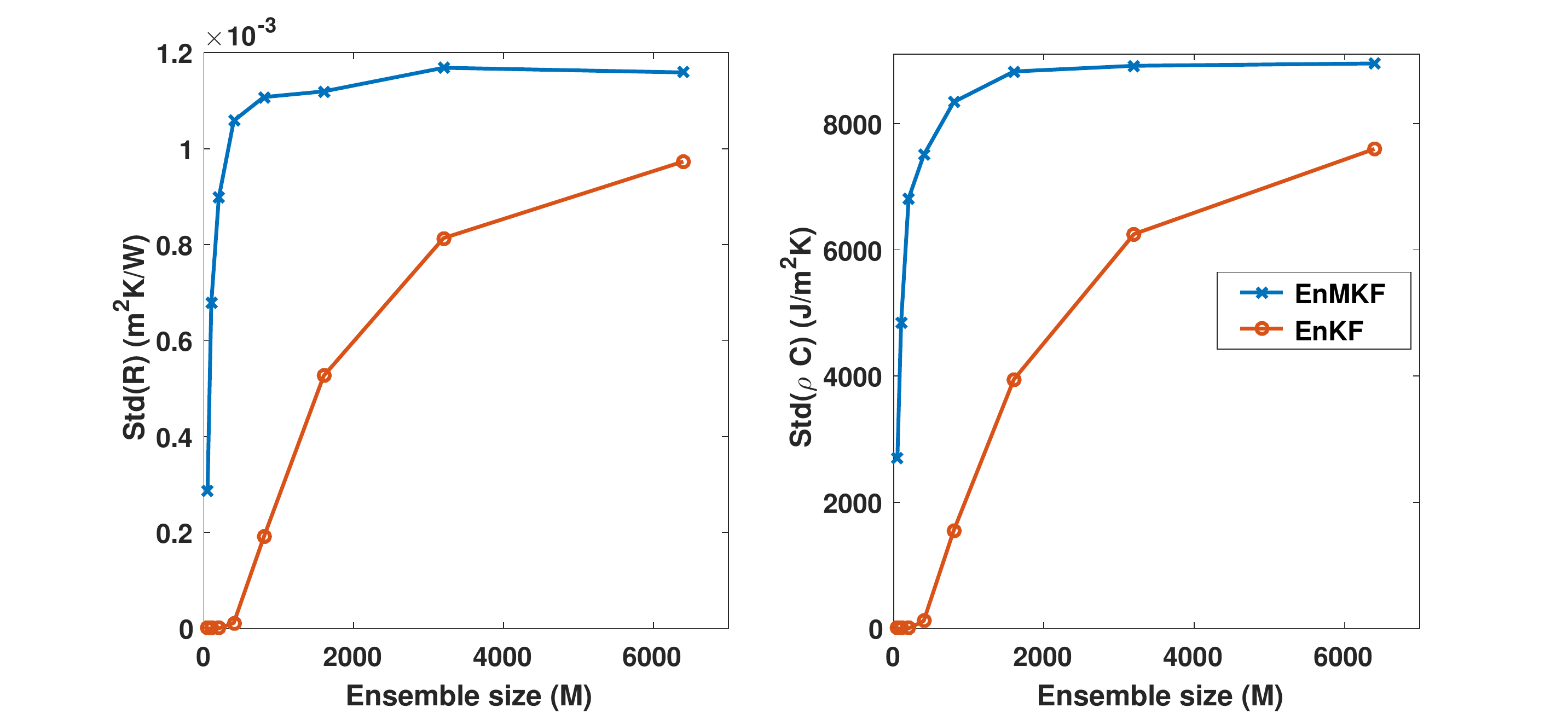}
\caption{Convergence of the ensemble standard deviation of the estimated parameters $R$ (left) and $\rho C$ (right) at time $t' = 3000$ with respect to the ensemble size $M$.}
\label{stdConv}
\end{figure}

Finally, we compare the EnMKF method and the modified EnKF method by analyzing the convergence of the estimated parameters at a fixed time with respect to the ensemble size $M$. Figures \ref{meanConv}, \ref{meanErrorConv}, and \ref{stdConv} show the convergence of the parameters-ensemble mean and standard deviation with ensemble size $M$ at time $t' = 3000$ minutes. The ensemble mean converges linearly for both methods but EnMKF has a better error constant and, therefore, is more reliable for small ensemble sizes. Moreover, due to the collapse of the ensemble, the convergence of the ensemble standard deviation for EnKF occurs only after $M>1000$. 

\subsection{Stopping criteria}
\label{converge}
EnMKF solves the collapse problem that usually arises with EnKF algorithms for state-parameter estimation. Ideally, we want the ensemble of each parameter to converge with time to become samples from the posterior distribution. Assuming this posterior distribution is unimodal, the convergence of our algorithm can be confirmed by computing the ensemble mean and standard deviation for each parameter. The time convergence of these estimated values can be used to suggest stopping criteria for ongoing measurement campaigns. For example, when the difference between two consecutive means and two consecutive standard deviations are relatively small, then the measurement campaign should be stopped.

\begin{figure}[h!]
\centering
\includegraphics[scale=0.6]{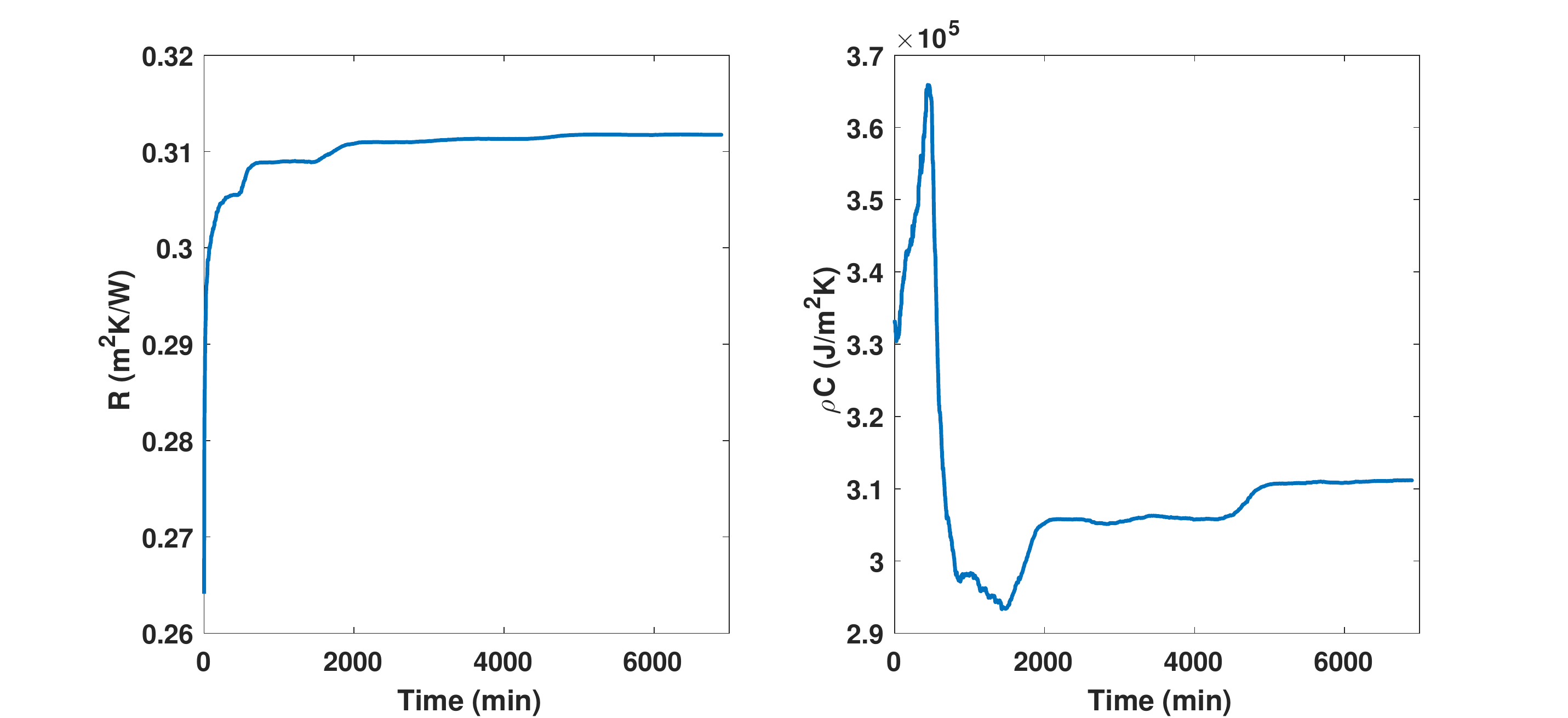}
\caption{Time convergence for the thermal resistance mean (left) and heat capacity mean (right) using EnMKF with ensemble size $M = 100$.}
\label{stopc1}
\end{figure}

\begin{figure}[h!]
\centering
\includegraphics[scale=0.6]{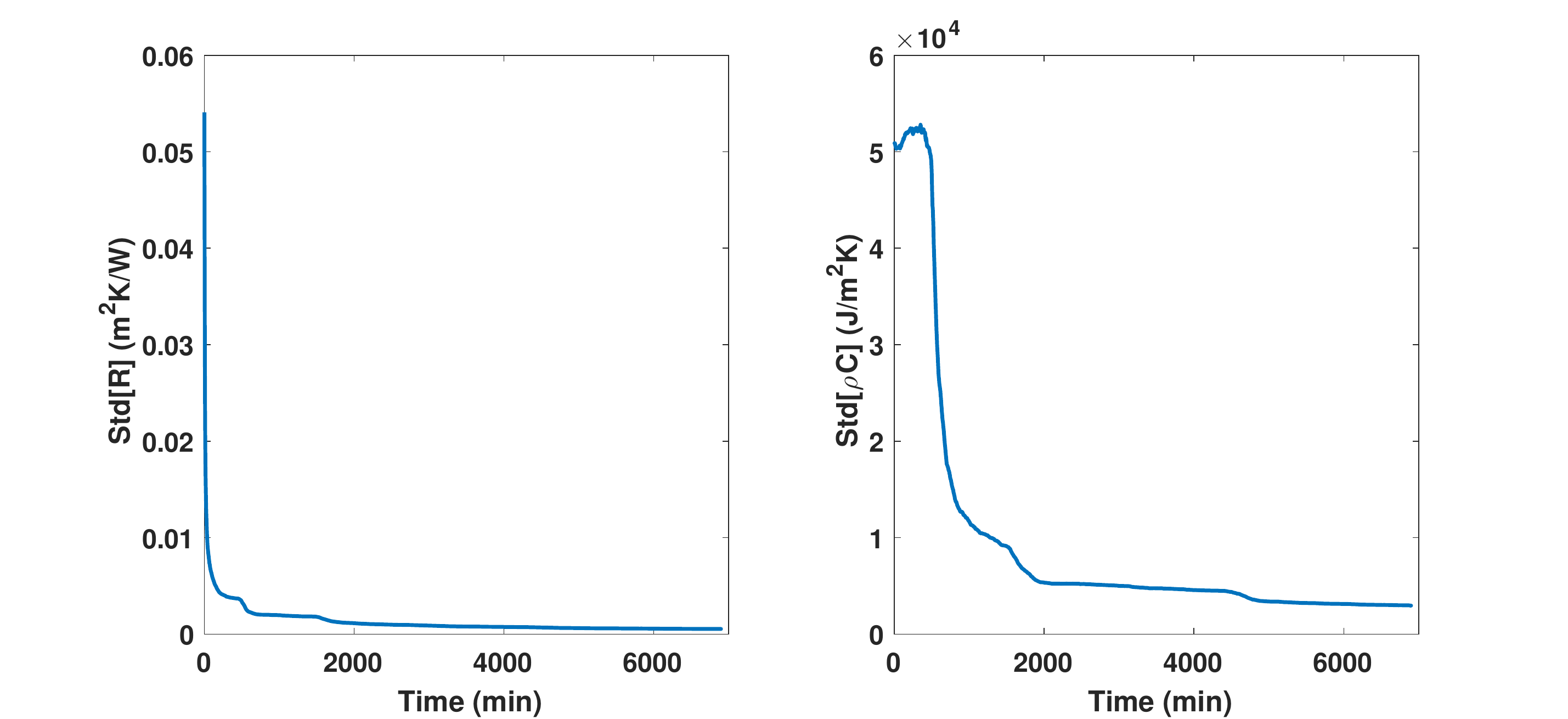}
\caption{Time convergence for the thermal resistance standard deviation (left) and heat capacity standard deviation (right) using EnMKF with ensemble size $M = 100$.}
\label{stopc2}
\end{figure}

In Figures \ref{stopc1} and \ref{stopc2}, we see that the ensemble mean and standard deviation for $R$ and $\rho C$ stabilize after $5000$ minutes. This indicates that the measurements collected up to that time are sufficient and the experiment can be stopped. 

\subsection{Bias error analysis}
\label{biaserror}
\begin{figure}[h!]
\centering
\includegraphics[width=0.8\textwidth]{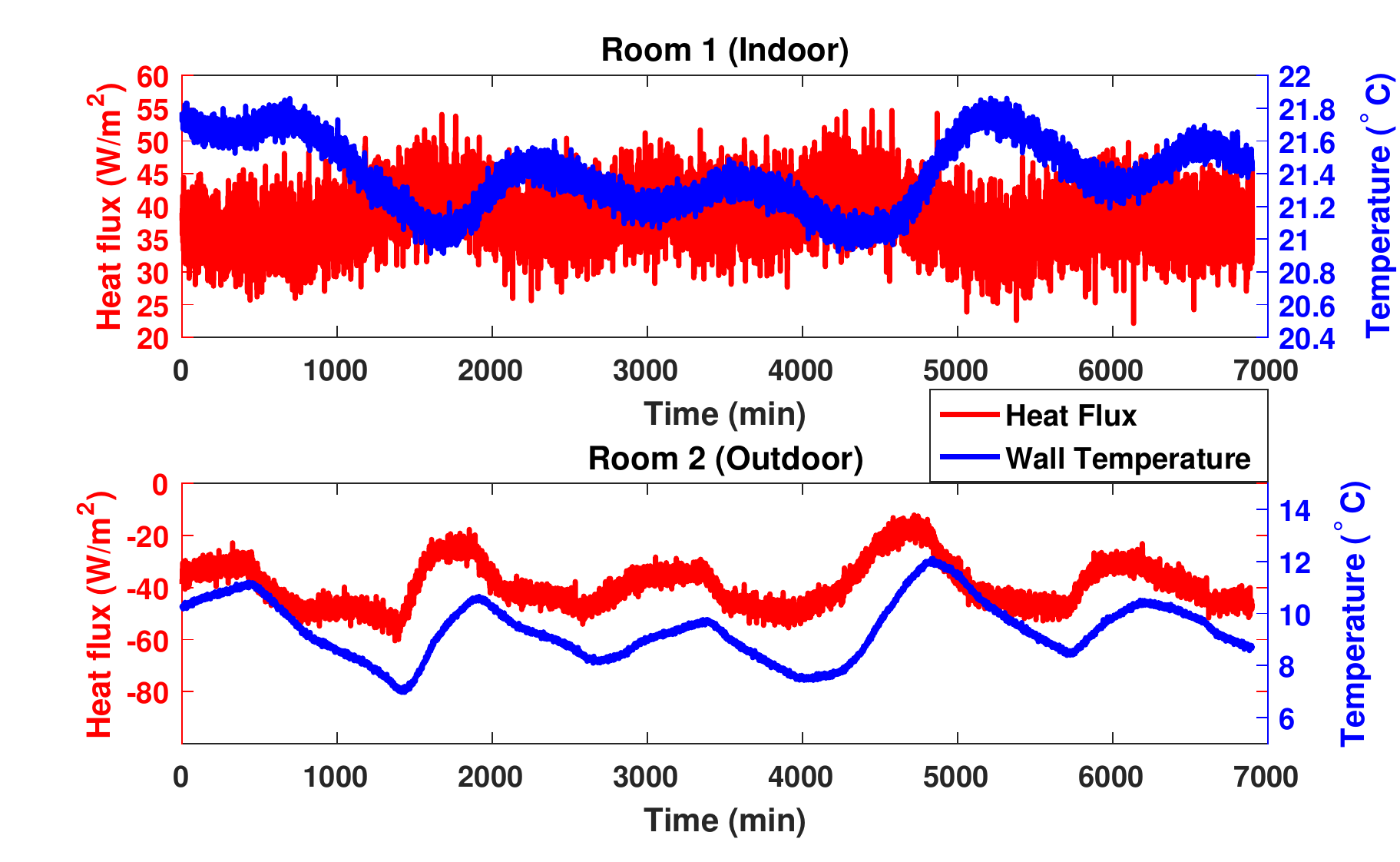}
\caption{Synthetic data measurements of temperature and heat flux generated using $R=0.3106$ and $\rho C = 3.2 \times 10^5$. }
\label{SyntheticData}
\end{figure}

We treat the boundary conditions as random variables that are marginalized in EnMKF and sampled in EnKF. In both EnKF and EnMKF, the bias error of the estimated parameters and the variability of the estimated state are reduced by increasing the ensemble size. However, in EnKF, the bias error induced by the boundary conditions depends on the ensemble size, $M$; with EnMKF, that bias error is eliminated independently from $M$. Therefore, EnMKF reduces the total bias error in comparison with the EnKF. To strengthen this conclusion, we use a synthetic data set similar to the experimental data set from Nottingham University (Figure \ref{fig1}). The synthetic data are generated by assuming $R = 0.3106\, m^2K/W$, $\rho C = 3.2\times10^5 J/m^2K$, smooth boundary conditions, and the initial condition given in \eqref{model_as}. We solve the heat equation and compute the heat flux at the boundaries for each minute. The temperature and heat-flux time series are then perturbed by Gaussian white noise. The resulting synthetic data are presented in Figure \ref{SyntheticData}.

We apply the Kalman filter to the perturbed temperature series and obtain the filtered mean temperatures and their corresponding variances. Then, we apply EnMKF and EnKF to estimate the parameters $R$ and $\rho C$ using the uniform priors: 

\[ R \sim U(0.28, 0.36) \, , \:\rho C \sim U(301000, 376000)\,. \] 

\begin{figure}[h!]
\centering
\includegraphics[scale=0.6]{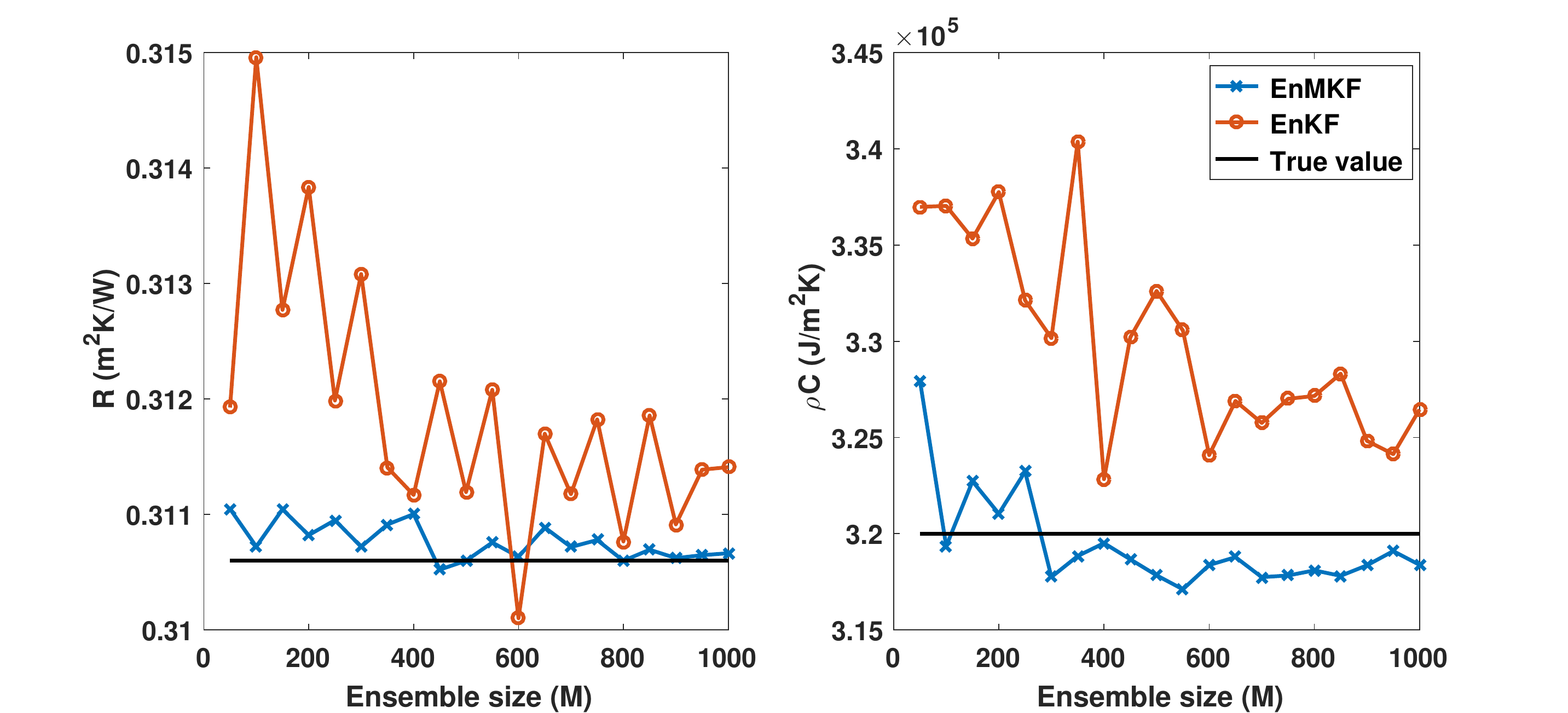}
\caption{Convergence of the ensemble mean of the estimated parameters $R$ (left) and $\rho C$ (right) at time $t' = 2000$ with respect to the ensemble size $M$.}
\label{meanConvSyn}
\end{figure}  

Figure \ref{meanConvSyn} shows the convergence of the sample mean of the estimated parameters using EnMKF and EnKF at time $2000$ with respect to the ensemble size. The total bias error obtained by EnMKF is smaller and more stable than the error produced by the modified EnKF. For large ensemble sizes, both methods provide similar results.


In this case, the synthetic data have Gaussian noise. In the previous sections, the real noise is not Gaussian. Nevertheless, the results show that the influence of Gaussianity assumption is not very significant when the sequential framework is applied to this heat-transfer problem.

\section{Summary and Conclusions}
\label{conc}

We studied the filtering problem for state observation system with random control vector and unknown static parameters. The control vectors may represent the boundary conditions or the source function of the underlying PDE model. By creating an artificial linear dynamic model for the control vector, we applied Kalman filter to obtain a Gaussian distribution of the random control vectors at each time step. The resulting distributions were then incorporated to generate a marginalized Kalman filter for the conditional state. The ensemble-marginalized Kalman filter (EnMKF) was derived using conditional probability, in particular, the law of total covariance. The main feature of EnMKF appeared in the prediction covariance which is approximated by the sample covariance of the augmented vector inflated by $E_{\Theta}\left[ Cov[B_{\theta} u_{k}] \right]$. Another sample-based EnKF algorithm was also proposed where the prediction covariance is only approximated by the sample covariance. 

To deal with high-dimensional joint state-parameter estimation problems, fully Bayesian techniques such as particle filters are not computationally affordable. The two introduced algorithms are not fully Bayesian; thus, they may handle nonlinear state-parameter systems and high-dimensional problems. EnKF usually requires additional adjustments in order to provide better approximations to the desired posterior distribution. EnMKF, on the other hand, completely avoids the ensemble-collapse phenomenon.

We applied both algorithms to estimate the thermal properties of a solid brick wall from experimental data. We showed that EnMKF provides better results than the modified EnKF. Due to the marginalization technique in EnMKF, we prevented the collapse of the parameters ensemble without artificial inflation. EnMKF had other advantages as well, such as a reduced heat-flux variance and more accurate fit of the measurements. We also considered a synthetic data example and compared the bias error from the parameters estimated using EnMKF and EnKF. The results confirmed that EnMKF provides a smaller bias error when the ensemble size is not very large. Finally, we proposed stopping criteria to reduce the time and cost required for real experiments, based on the estimated parameters reaching a stationary period.

While the numerical results are focused on the heat-transfer application, the EnMKF can be used for a wider PDE-constrained data assimilation problems in which the boundary conditions are subject to uncertainty; these include, for example, flood predictions via hydrological models \cite{PAPPENBERGER20061430}.

\section*{Acknowledgements}
Z. Sawlan, M. Scavino and R. Tempone are members of the KAUST SRI Center for Uncertainty Quantification in Computational Science and Engineering. R. Tempone received support from the KAUST CRG3 Award Ref: 2281 and the KAUST CRG4 Award Ref: 2584.


\bibliography{mybib_new}

\end{document}